\documentclass[reprint,prd,twocolumn,superscriptaddress,nofootinbib]{revtex4-2}   
\usepackage{graphicx}
\usepackage{multirow}
\usepackage{gensymb}
\usepackage{amsmath}
\usepackage{bm}
\usepackage{natbib}
\usepackage{hyperref}
\usepackage{color}  
\newcommand\araa{{ARA\&A}}
\renewcommand\apj{{ApJ}}
\newcommand\apjl{{ApJL}}
\newcommand\apjs{{ApJS}}
\newcommand\aap{{A\&A}}
\newcommand\aapr{{A\&A~Rev.}}
\newcommand\mnras{{MNRAS}}
\renewcommand\prd{{Phys.~Rev.~D}}
\newcommand\pasp{{PASP}}
\renewcommand\nat{{Nature}}
\newcommand\pasa{{PASA}}

\def\mone{{\rm M}_1}

\def\nsampi{{\rm N}_{\rm samp,I}}
\def\nsampymc{{\rm N}_{\rm samp,YMC}}
\def\nsampiso{{\rm N}_{\rm samp,IB}}

\def\msampi{{\rm M}_{\rm samp,I}}
\def\msampymc{{\rm M}_{\rm samp,YMC}}
\def\msampiso{{\rm M}_{\rm samp,IB}}

\def\nmrgi{{\rm N}_{\rm mrg,I}}
\def\rsfci{\zeta_{\rm 0,I}}

\def\zmax{z_{\rm max}}

\def\zf{z_{\rm f}}
\def\tf{t_{\rm f}}

\def\tevnt{t_{\rm event}}
\def\zevnt{z_{\rm event}}

\def\tld{t_{\rm lD}}
\def\tobs{t_{\rm obs}}
\def\delobs{\Delta t_{\rm obs}}
\def\delobsi{\Delta t_{\rm obs,I}}
\def\delage{\Delta t_{\rm age}}

\def\rate{\mathcal R}
\def\ratei{{\mathcal R}_{\rm I}}

\def\rz{{\mathcal R}^{\prime}}
\def\rzi{{\mathcal R}^{\prime}_{\rm I}}
\def\rzymc{{\mathcal R}^{\prime}_{\rm YMC}}
\def\rzib{{\mathcal R}^{\prime}_{\rm IB}}

\def\rymc{{\mathcal R}_{\rm YMC}}
\def\riso{{\mathcal R}_{\rm IB}}

\def\clmf{\phi_{\rm CLMF}}
\def\sfh{\Phi_{\rm SFH}}

\def\peryg{{\rm~yr}^{-1}{\rm Gpc}^{-3}}

\def\etaymc{\eta_{\rm YMC}}
\def\etaiso{\eta_{\rm IB}}
\def\fimf{f_{\rm IMF}}
\def\mupi{\mu_{\rm I}^p}
\def\mupymc{\mu_{\rm YMC}^p}
\def\mupiso{\mu_{\rm IB}^p}
\def\muobs{{\bm \mu}_{\rm obs}^{\bm p}}
\def\muobsi{\mu_{\rm obs,i}^p}
\def\mup{\mu^p}
\def\norm{{\mathcal N}}
\def\lhood{{\mathcal L}}
\def\nobs{N_{\rm obs}}
\def\unif{{\mathcal U}}

\newcommand{\Ms}{\ensuremath{{\rm M}_{\odot}}}

\newcommand{\eg}{{\it e.g.}}

\newcommand{\ie}{{\it i.e.}}

\newcommand{\beq}{\begin{equation}}
\newcommand{\eeq}{\end{equation}}

\newcommand{\kmps}{\ensuremath{{\rm~km~s}^{-1}}}
\newcommand{\peryr}{\ensuremath{{\rm~yr}^{-1}}}

\newcommand{\thub}{\ensuremath{t_{\rm Hubble}}}

\newcommand{\mcl}{\ensuremath{M_{\rm cl}}}
\newcommand{\rh}{\ensuremath{r_{\rm h}}}

\newcommand{\tmrg}{\ensuremath{t_{\rm mrg}}}

\newcommand{\nbseven}{{\tt NBODY7}}

\newcommand{\bse}{{\tt BSE}}

\newcommand{\archain}{{\tt ARCHAIN}}

\newcommand{\pymc}{{\tt PyMC3}}

\newcommand{\fbin}{\ensuremath{f_{\rm bin}}}

\newcommand{\fobin}{\ensuremath{f_{\rm Obin}}}
\newcommand{\fymc}{\ensuremath{f_{\rm YMC}}}

\newcommand{\ace}{\ensuremath{\alpha_{\rm CE}}}

\begin{document}

\title[BBH merger rate density from multiple channels]
{Merger rate density of stellar-mass binary black holes from young massive clusters,
open clusters, and isolated binaries: comparisons with LIGO-Virgo-KAGRA results}

\author{Sambaran Banerjee}
\email{sambaran.banerjee@gmail.com (he/him/his)}
\affiliation{Helmholtz-Instituts f\"ur Strahlen- und Kernphysik (HISKP),
Nussallee 14-16, D-53115 Bonn, Germany}
\affiliation{Argelander-Institut f\"ur Astronomie (AIfA), Auf dem H\"ugel 71,
D-53121, Bonn, Germany}

\date{\today}

\begin{abstract}
I investigate the roles of cluster dynamics and massive binary evolution in
producing stellar-remnant binary black hole (BBH) mergers over the cosmic time.
To that end, dynamical BBH mergers are obtained from long-term direct N-body evolutionary models of
$\sim10^4\Ms$, pc-scale young massive clusters (YMC) evolving into moderate-mass open clusters
(OC). Fast evolutionary models of massive isolated binaries (IB) yield BBHs from
binary evolution. Population synthesis in a Model Universe is then performed, taking into account observed
cosmic star-formation and enrichment histories, to obtain BBH-merger yields from
these two channels observable at the present day and over cosmic time. The merging
BBH populations from the two channels are combined by applying a proof-of-concept Bayesian
regression chain, taking into account observed differential intrinsic BBH merger rate densities
from the second gravitational-wave transient catalogue (GWTC-2). The analysis estimates
an OB-star binary fraction of $\fobin\gtrsim90$\% and a YMC formation efficiency of
$\fymc\sim10^{-2}$, being consistent with recent optical observations and large scale structure
formation simulations. The corresponding combined Model Universe present-day,
differential intrinsic BBH merger rate density and the cosmic evolution of
BBH merger rate density both agree well with those from GWTC-2. The analysis
also suggests that despite significant `dynamical mixing' at low redshifts,
BBH mergers at high redshifts ($\zevnt\gtrsim1$) could still be predominantly determined by
binary-evolution physics. Caveats in the present approach and future improvements are discussed. 
\end{abstract}

\maketitle

\section{Introduction}\label{intro}

We are approaching a golden era of
detections of binary stellar remnant (or compact binary) merger events in gravitational waves (hereafter GW)
and of multi-messenger astronomy. Such events,
which are mergers of binaries comprising of neutron stars (hereafter NS)
and stellar-remnant black holes (hereafter BH), are among the most energetic transient events in
the Universe in GW and, potentially, in electromagnetic waves.
Recently,
the LIGO-Virgo-KAGRA collaboration (hereafter LVK)\citep{Asai_2015,Acernese_2015,KAGRA_2020}
has published,
in their second GW transient catalogue (hereafter GWTC-2)\citep{Abbott_GWTC2,Abbott_GWTC2_prop},
47 candidates
(false alarm rate of $<1\peryr$) of compact binary merger events from
until the first half, `O3a', of their third observing run, `O3'. GWTC-2
includes all GWTC-1 events \citep{Abbott_GWTC1,Abbott_GWTC1_prop}
from the previous LIGO-Virgo `O1' and `O2' observing runs.
Based on the parameter estimations of these events, the vast majority of 
them have been designated as binary black hole (hereafter BBH) mergers with
component masses ranging through $\approx5\Ms-100\Ms$ \citep{Abbott_GWTC2_prop}.
The rest comprise candidates of binary neutron star (hereafter BNS),
neutron star-black hole (hereafter NS-BH), and `mass-gap' \citep{Abbott_GW190814} mergers.
Additional candidate events from
the second half of O3 are just being announced \citep{Abbott_NSBH_2021,Abbott_GWTC2.1}.

The plethora of observed GW events have naturally triggered exploration of
a wide range of theoretical scenarios or `channels' that model pairing of NSs and BHs and
their approach towards general relativistic (hereafter GR) inspiral
and merger. The various channels can be broadly
classified as `dynamical' and `isolated binary evolution' channels
\citep{Benacquista_2013,Mandel_2017,Mapelli_2018}. The dynamical channels involve
pairings and mergers mediated by dynamical interactions in dense stellar
systems such as young clusters, open clusters, globular clusters (hereafter GC),
nuclear clusters \citep[\eg,][]{DiCarlo_2019,Banerjee_2017,Kumamoto_2020,Askar_2016,Kremer_2020,Hoang_2017}
and in hierarchical or chaotic systems in galactic fields \citep[\eg,][]{Antonini_2017a,Yu_2020,Fragione_2020,Samsing_2014,Michaely_2019}.
In isolated-binary channels, galactic-field binaries comprising progenitor stars
of NSs and BHs directly hatch, through binary evolution and
without involvement in dynamical interactions, compact binaries
tight enough to merge within a Hubble time
\citep[\eg,][]{Dominik_2012,Belczynski_2016,DeMink_2016,Marchant_2016,Stevenson_2017,Giacobbo_2018,Kruckow_2018,Breivik_2020,Bavera_2020}. Merger channels can also be `hybrid' in the sense that
both binary evolution and dynamical interaction in clusters or the field
play role in assembling the compact binary and driving its merger
\citep[\eg,][]{Gonzalez_2020,VignaGomez_2021,Hamers_2021}.
Another hybrid channel is the interplay between hydrodynamic drag and dynamical
interactions in, \eg, gas discs of active galactic nuclei \citep{McKernan_2018,Secunda_2019}. 
However, the current GW observations do not rule out or prefer any particular channel(s)
over others and it is quite likely that multiple channels contribute significantly
to the observed GW events, given the wide landscape
of these events and the several unknown/tunable parameters in the models for each channel
\citep{Zevin_2020}. This would hold true despite an individual
(sub-)channel may, over certain regions of its parameter space,
well reproduce one or more aspects of the observed event population (\eg, mass
distribution, rates; \cite{Santoliquido_2020,Rodriguez_2021,Banerjee_2021}). 

In this study, two such intriguing and well-explored BBH-merger channels are considered.
One is the dynamical interactions in star clusters that
`begin life' as young massive clusters (hereafter YMC)\citep{PortegiesZwart_2010}
and evolve into moderately-massive to massive open clusters (hereafter OC).
In the YMC phase ($\lesssim10$ Myr age of the bulk stellar population),
such clusters are, typically, observed to be
gas free, near spherical, of $\sim10^4\Ms-\sim10^5\Ms$,
and of $\sim$ pc length scale (viral radius). The BHs retained in these
clusters would continue to remain dynamically active in the
cluster's innermost region ($<<1$ pc) for at least several
Gyr, producing dynamically-assembled BBH mergers \citep{Banerjee_2010,Banerjee_2017}.
In this work, this channel will hereafter be referred to as the
YMC/OC channel. The other channel is the isolated binary (hereafter IB)
evolution (see above) - the IB channel. IBs having both components of
zero age main sequence (hereafter ZAMS) mass $\gtrsim5\Ms$ (depending on
metallicity) evolve into BNS, NS-BH, or BBH, depending on the component
masses and evolutionary history \citep{Tutukov_1979,Hurley_2002}.

Here, a proof-of-concept linear Bayesian regression chain is applied
to combine the BBH-merger outcomes from model YMC/OC and IB populations. 
The regression is performed based on the present-day, differential intrinsic
BBH merger rate densities estimated from GWTC-2 \citep{Abbott_GWTC2_prop}.
Sec.~\ref{ymcmodel} and \ref{isomodel} describe computations of evolutionary models of
YMC/OC and IB, respectively. Sec.~\ref{popsynth} describes
cosmological population synthesis of BBH mergers based on
outcomes from these evolutionary models.
Sec.~\ref{res} describes the Bayesian regression
for combining the outcomes from the YMC/OC and IB populations and demonstrates
comparisons with GWTC-2 BBH merger rates: both, the present-day
differential rates and the cosmic rate evolution. Sec.~\ref{summary}
summarizes the results and discusses caveats and future developments.

\section{Computations: cosmological population synthesis of
star clusters and isolated field binaries}\label{comp}

\subsection{Many-body, relativistic, evolutionary models of
young massive and open star clusters}\label{ymcmodel}

In this work, the long-term evolutionary model set of YMCs/OCs as described in
Ref.~\cite{Banerjee_2021} is utilized. The various model ingredients, the
computational approach, and their astrophysical
implications are described in detail in
Refs.~\cite{Banerjee_2020,Banerjee_2020c,Banerjee_2020d,Banerjee_2021}. Therefore,
only a summary of these computations is presented here. 

The model star clusters, initially,
have masses of $2\times10^4\Ms\leq\mcl\leq10^5\Ms$ and sizes (half-mass radii) of 
$1{\rm~pc}\leq\rh\leq2{\rm~pc}$. Their metallicities range over 
$0.0001\leq Z \leq0.02$ and they orbit in a solar-neighborhood-like
external galactic field. The initial models are composed of
ZAMS stars of masses $0.08\Ms\leq m_\ast\leq150.0\Ms$
that are distributed according
to the Kroupa initial mass function (hereafter IMF) \citep{Kroupa_2001}, $\fimf(m_\ast)$.
About half of the models
have a primordial-binary population (overall initial binary fraction
$\fbin\approx5$\% or 10\%) where all O-type stars (\ie, stars
with $m_\ast\geq16\Ms$) are paired among themselves
(\ie, initial binary fraction among O-stars is $\fobin=100$\%)
according to an observationally-deduced distribution of massive-star binaries
\citep{Sana_2011,Sana_2013,Moe_2017}. Such cluster
parameters and stellar compositions are consistent with those
observed in `fully'-assembled, (near-)spherical, (near-)gas-free YMCs and
medium-mass OCs \citep{PortegiesZwart_2010,Banerjee_2017c,Banerjee_2018b,Krumholz_2019}
that continue to form, evolve, and dissolve in the Milky Way and other galaxies
(as such, anywhere in the Universe) active in star formation.

These model clusters are realistically evolved due to two-body relaxation \citep{Spitzer_1987},
close (relativistic) dynamical encounters \citep{Heggie_2003}
(without applying any gravitational softening),
and stellar evolution \citep{Pols_1998,Kippenhahn_2012}. This is achieved 
using the $\nbseven$ code,
a state-of-the-art post-Newtonian (hereafter PN) direct N-body integrator
\citep{Aarseth_2003,Aarseth_2012,Nitadori_2012}, that couples with the
semi-analytical stellar and binary-evolutionary scheme
$\bse$ \citep{Hurley_2000,Hurley_2002}. The integrated $\bse$ is made
up to date, in regards to prescriptions of stellar wind mass loss
and formation of NSs and BHs,
as detailed in Ref.~\cite{Banerjee_2020}.
NSs and BHs form according to
the `rapid' or `delayed' core-collapse supernova (hereafter SN) models of Ref.~\cite{Fryer_2012}
\footnote{The majority of the computed models of \cite{Banerjee_2021}
employ the rapid-SN prescription although a few models employ the delayed-SN prescription,
for exploratory purposes. The dynamical evolution and GR-merger outcomes
of the clusters are unlikely to be significantly
affected by this difference, as discussed
in Refs.~\cite{Banerjee_2020,Banerjee_2020d}.}
and
pulsation pair-instability SN (PPSN) and pair-instability SN (PSN) 
models of Ref.~\cite{Belczynski_2016a}. A newly formed NS or BH
receives natal kick that is modulated based on SN fallback onto it,
as in Ref.~\cite{Belczynski_2008}. Due to conservation of linear momentum,
such material fallback slows down the remnants, causing BHs of $\gtrsim10\Ms$
to retain in the clusters right after their birth. The material fallback
also plays role in shaping the mass distribution of NSs and BHs.
Furthermore, NSs formed via electron-capture SN (hereafter ECS) \citep{Podsiadlowski_2004}
also receive small natal kicks (of a few $\kmps$) and are retained in the clusters
at birth \citep{Gessner_2018}. See Ref.~\cite{Banerjee_2020} for further
details.

In $\nbseven$, the PN treatment is handled by the $\archain$ algorithm
\citep{Mikkola_1999,Mikkola_2008}. Such a PN treatment allows for
GR evolution of the innermost NS- and/or BH-containing binary
of an in-cluster (\ie, gravitationally bound to the cluster)
triple or higher order compact subsystem, in tandem with the Newtonian-dynamical
evolution of the subsystem (Kozai-Lidov oscillation or
chaotic three-body interaction), potentially leading
to the binary's (in-cluster) GR in-spiral and merger. The PN treatment
applies also to the GR evolution of in-cluster NS/BH-containing binaries that
are not a part of a higher-order subsystem. As discussed in previous
studies \citep{Banerjee_2010,Banerjee_2017,Banerjee_2018,Banerjee_2020c,Anagnostou_2020},
the moderate density and velocity dispersion
in the model clusters make them efficient in dynamically
assembling PN subsystems, particularly, those comprising BHs.
This causes the vast majority of the GR mergers
from these computed clusters to be in-cluster BBH mergers.
As also recently demonstrated \citep{Banerjee_2020d},
the final in-spiralling phase of such merging BBHs
sweep through the LISA and deci-Hertz GW frequency bands
before merging in the LVK band.

The model grid used in this work comprises 64 long term ($\sim10$ Gyr) evolutionary
cluster models (see Table~A1 of Ref.~\cite{Banerjee_2021}). 

\subsection{Evolutionary models of isolated binary populations}\label{isomodel}

To obtain an IB counterpart of the YMC/OC's dynamical BBH mergers,
populations of stellar binaries are evolved. This is done utilizing a
standalone version of the same $\bse$ that is coupled with $\nbseven$ \citep{Banerjee_2020}.
This standalone $\bse$ incorporates exactly
the same astrophysical ingredients and their implementations as in
$\nbseven/\bse$ (see Sec.~\ref{ymcmodel}).
Note that this updated $\bse$
preserves the original binary-evolution physics of Ref.~\cite{Hurley_2002},
except that the recipes for assigning masses of NSs and BHs and their natal kicks are updated.
In particular, the `$\alpha-\lambda$' prescription
\citep{Tutukov_1979,Ivanova_2013,Toonen_2016} is applied
for treating the common envelope (hereafter CE) evolution which process
is crucial and dominant for tight, merging double compact binary formation.
(In contrast, recent studies \cite[\eg,][]{Marchant_2021,Gallegos_2021}
involving binary evolution with one-dimensional hydro code
suggest stable mass transfer as the dominant channel for merging BBH production.)
A similar approach has been followed in other recent, independent studies
\citep{Giacobbo_2018,Baibhav_2019,Santoliquido_2020}.
Also, an `optimistic' scenario \citep{Riley_2021} for Hertzsprung-gap (hereafter HG)
stars is assumed as in Ref.~\cite{Hurley_2002} where HG donors are allowed to survive the
CE phase (as opposed to in, \eg, Refs.~\cite{Giacobbo_2018,Riley_2021}).

As in the YMC/OC models, the
distributions of semi-major-axis and eccentricity of the member binaries
of the model binary population
follow those of Ref.~\cite{Sana_2011}. The ZAMS masses of the binary components
are drawn from the standard IMF with $m_\ast\geq5\Ms$ and are paired randomly. 
In this way, a population comprising of $10^6$ binaries is generated \citep{Kuepper_2011}. 
The binaries are then evolved (individually, one by one
\footnote{Since $\bse$ is a semi-analytic code the standalone $\bse$ runs
can be computed easily, despite the large number of binaries. That way,
with only a moderate computational cost, good statistics can be obtained for
the $\bse$ runs.}
) with the standalone $\bse$. The $\bse$ evolutions are performed for
metallicities $Z=0.0001$, 0.0002, 0.001, 0.005, 0.01, and 0.02
and for CE efficiency parameters \citep{Ivanova_2013}
$\ace=1.0$, and 3.0 (\ie, a total of 12 evolutionary sets of the $10^6$ binaries).
In all the $\bse$ runs, the `rapid' remnant mass scheme along with PPSN/PSN and
ECS-NS formation \citep{Banerjee_2020} is applied.
The natal kicks of all NSs and BHs formed during the binary evolution
are moderated due to SN material fallback according to the
conservation of linear momentum (the `momentum-conserving' natal kick \cite{Belczynski_2008,Banerjee_2020}).
The unmoderated natal
kicks of core-collapse SN remnants are distributed according to
a Maxwellian with one-dimensional dispersion of $\sigma_{\rm CC}=265\kmps$ \citep{Hobbs_2005}.
ECS-NSs, on the other hand, receive much lower natal kicks
of one-dimensional dispersion $\sigma_{\rm ECS}=3\kmps$ \citep{Gessner_2018}.
Note that these same values and natal-kick prescription are applied also in the N-body models of YMC/OCs
determining the retention of BHs and NSs in the clusters (Sec.~\ref{ymcmodel}).
All mass-transfer
episodes are treated with Eddington-factor limited accretion onto the recipient member
\citep{Hurley_2002}.

A fraction of the binaries evolve into double-compact (\ie, BBH, BNS, NS-BH)
binaries as a result of the $\bse$ binary-evolutionary scheme. The GR inspiral and
merger of these binaries are tracked by simply applying the orbit-averaged
quadrapole GW radiation formulae \citep{Peters_1964}.
The double-compact binaries
that merge within the Hubble time are, typically, survivors of CE evolution
and/or mass-transfer phases
\citep{Belczynski_2002,Belczynski_2016,Mandel_2017,Stevenson_2017,Giacobbo_2018,Chatto_2021}.
Since the vast majority of
such double-compact binaries have small or zero eccentricity at formation
(unlike the dynamically-assembled/triggered merging binaries), the
orbit-averaged treatment of the GR inspiral serves as a reasonable approximation.

\subsection{Cosmological population synthesis of star clusters and
isolated binaries}\label{popsynth}

To estimate the BBH merger rate density (both present-day and at higher redshifts) 
from the evolutionary YMC/OC and IB model grids, a Model Universe is constructed
comprising of YMC/OCs or IBs or a combination of these, following the same
approach as described in Ref.~\cite{Banerjee_2021}.
In such a Model Universe,
a YMC or a burst of IB population is formed at a redshift $\zf$, that
corresponds to the age of the Universe $\tf$. $\zf$ is taken to be distributed
according to the observed cosmic star formation history (hereafter SFH) as given by \citep{Madau_2017}
\beq
\sfh(\zf) = 0.01\frac{(1+\zf)^{2.6}}{1+[(1+\zf)/3.2]^{6.2}} \Ms{\rm~yr}^{-1}{\rm~Mpc}^{-3}.
\label{eq:mdsfr}
\eeq

The YMCs and/or the IB-bursts are assumed to be uniformly distributed
within an effective detector visibility horizon at redshift $\zmax$ \citep{Chen_2017b}
and contribute to the present-day, observed in-spiral/merger events.
A GR merger occurs
from a parent stellar population (a YMC or an IB-population)
$\tmrg$ `delay time' after the population's birth,
when the age of the Universe is $\tevnt$ (corresponding to a redshift $\zevnt$), \ie,
\beq
\tevnt = \tf + \tmrg. 
\label{eq:tevnt}
\eeq
If the light travel time from the population's comoving (or Hubble) distance, $D$, is $\tld$ then
the age of the Universe is
\beq
\tobs = \tevnt + \tld
\label{eq:tobs}
\eeq
when the (redshifted) GW signal from the merger event arrives the detector.
The GW signal is considered `present-day' (or `recent' or `in the present epoch') if
\beq
\thub-\delobs \leq \tobs \leq \thub+\delobs
\label{eq:delobs}
\eeq
where $\thub$ is the current age of the Universe (the Hubble time) and $\delobs$ is a
tolerance time interval. $\delobs$ serves as an uncertainty in the formation
epoch of the parent stellar population which is $<1$ Gyr \citep{Madau_2014}.

In this work, hypothetical Model Universes are constructed by assuming that
the entire star formation of the Universe occurs in the form of YMCs or IBs. 
The resulting merger rate densities are then scaled or combined based on a
Bayesian linear regression analysis as described further below. The present-day
Model Universe merger events are obtained based on a sample
population of $\nsampi=10\times10^5$ (10 independent samples, each of $10^5$ members) YMC/OCs or
IB-populations spread uniformly within $\zmax$. From the computed YMC/OC evolutionary
model grid (see Sec.~\ref{ymcmodel}), the Model Universe members are randomly chosen with initial
masses according to a power-law of index -2 (\ie, $\clmf(\mcl)\propto \mcl^{\alpha}$; $\alpha=-2$)
as observations of young clusters in the Milky Way and
nearby galaxies suggest \citep{Gieles_2006b,Larsen_2009,PortegiesZwart_2010,Bastian_2012}.
Their initial sizes are chosen uniformly between $1{\rm~pc}\leq\rh\leq2{\rm~pc}$.
The IB-populated universe is analogously filled with the evolutionary model IB
populations (which always begin with $10^6$ binaries; see Sec.~\ref{isomodel}).
The metallicities of both, the clusters and the IBs, are chosen based on
the observation-based redshift-metallicity lookup tables of
Ref.~\citep{Chruslinska_2019}, in the same way as described in
Ref.~\citep{Banerjee_2021}. The present-day time tolerance is taken to be
$\delobs=0.15{\rm~Gyr}$ ($\delobs=0.005{\rm~Gyr}$)
\footnote{The much shorter $\delobs$ for the IB-universe is to avoid
an excessive number of present-day mergers (and hence a large volume of data to be handled)
in the Model Universe population
synthesis and make it comparable to that from the YMC/OC-universe. The IB-population,
without corrections (see below), produces a much larger number of mergers
per unit mass than that from YMC/OCs since, unlike the latter, the IB population
is `zoomed in' to $m_\ast\geq5\Ms$ (see Sec.~\ref{isomodel}).}
for the YMC/OC-filled (IB-filled) universe. The detector horizon is taken to be
$\zmax=1.0$ as applicable for LVK O3 \citep{Abbott_GWTC2}.

Let the total number of present-day merger events is $\nmrgi$, as obtained from the sample of
parent stellar population of type I (I$=$ YMC or IB) of total mass at birth
$\msampi$. Then the corresponding present-day Model Universe merger rate, per unit mass
of star formation (or present-day `specific merger rate'), is
\beq
\rsfci = \frac{\nmrgi}{(2\delobsi)\msampi}.
\label{eq:sfcrate}
\eeq
For the cluster-filled universe, $\msampymc$ is simply the sum of the
initial masses of the clusters in the sample population, \ie,
\beq
\msampymc = \sum_{i=1}^{\nsampymc}{\rm M}_{{\rm cl},i}.
\label{eq:mymc}
\eeq
For the IB-filled universe, due to the lower truncation of the ZAMS mass distribution
at $5\Ms$ (see Sec.~\ref{isomodel}), a corrective scaling to the total initial mass, ${\rm M}_{\rm IB}$,
of the $10^6$ binaries has to be applied, to account for the 
full standard-IMF over $0.08\Ms\leq m_\ast\leq150.0\Ms$ (as taken for the clusters).
Thus,
\beq
\msampiso = \frac{1}{f_\ast}\nsampiso{\rm M}_{\rm IB}  
\label{eq:miso}
\eeq
where
$f_\ast=(\int_{5.0}^{150}\fimf(m_\ast)dm_\ast)/
	      (\int_{0.08}^{150}\fimf(m_\ast)dm_\ast)$.

Note that $\rsfci$ already incorporates cosmic star formation and metallicity
evolution histories, merger delay time, and light travel time (see above). Therefore,
the present-day intrinsic merger rate density can be obtained by simply scaling
$\rsfci$ with the integrated star formation rate (hereafter SFR) as
\beq
\ratei = \rsfci\int_{t(z=10)}^{t(z=0)}\sfh(z(t))dt.
\label{eq:rate}
\eeq
Note that this approach corresponds to essentially performing the standard integral
over redshift, metallicity, and cosmic volume, for merger rate density calculation
(\eg, Eqn.~1 of Ref.~\cite{Santoliquido_2020}),
in a Monte Carlo fashion. The present-day merger events can be
binned against a merger property (\eg, primary mass, mass ratio), $X$.
The resulting normalized density function can then be scaled
by $\ratei$ to obtain the present-day intrinsic differential merger rate density as
\beq
\frac{d\ratei}{dX}(X)=\ratei\frac{1}{\nmrgi}\frac{d\nmrgi}{dX}(X).
\label{eq:diffrate}
\eeq
Here,
\beq
\frac{d\nmrgi}{dX}(X) \approx \left.\frac{\Delta\nmrgi}{\Delta X}\right|_X
\label{eq:ratepdf}
\eeq
where $\Delta\nmrgi$ is the event count over a bin of width $\Delta X$ 
around the value $X$. In this study, 40 bins over $5\Ms\leq\mone\leq85\Ms$
and 20 bins over $0.1\leq q\leq1.0$ are used to construct
differential merger rate densities.

\begin{figure*}
\centering
\textbf{Pure channel: clusters (YMC/OC) and isolated binaries (IB; $\ace=1$ and $3$)}\par\medskip
\includegraphics[width=15.0cm,angle=0]{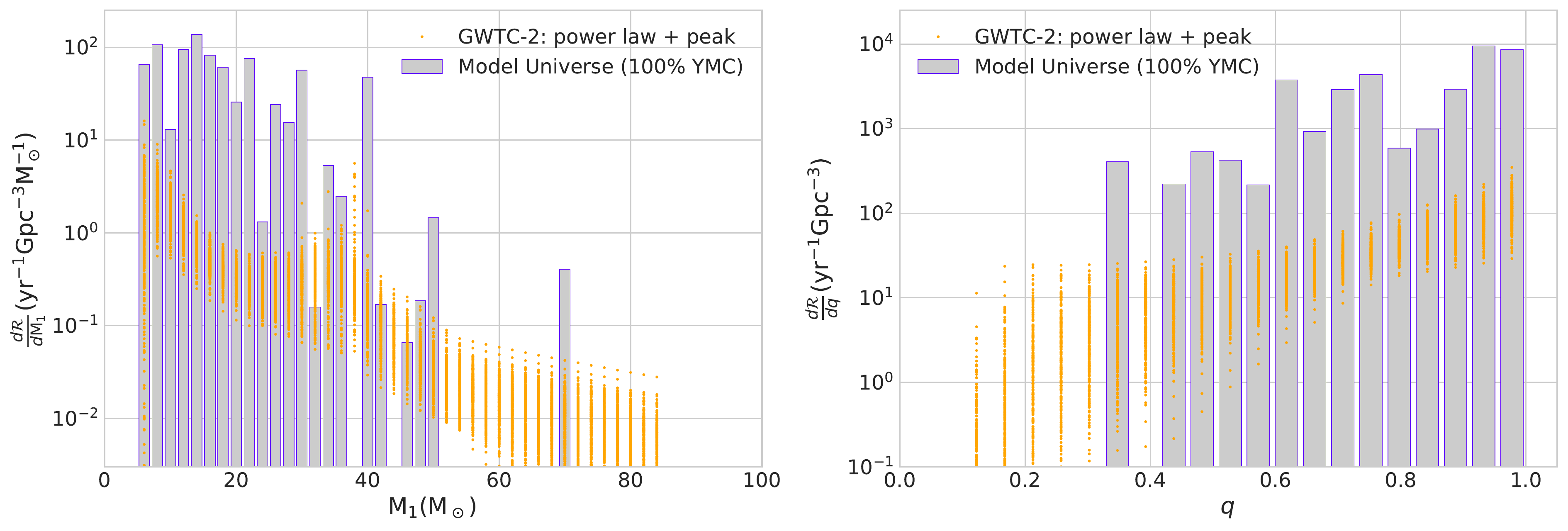}\\
\includegraphics[width=15.0cm,angle=0]{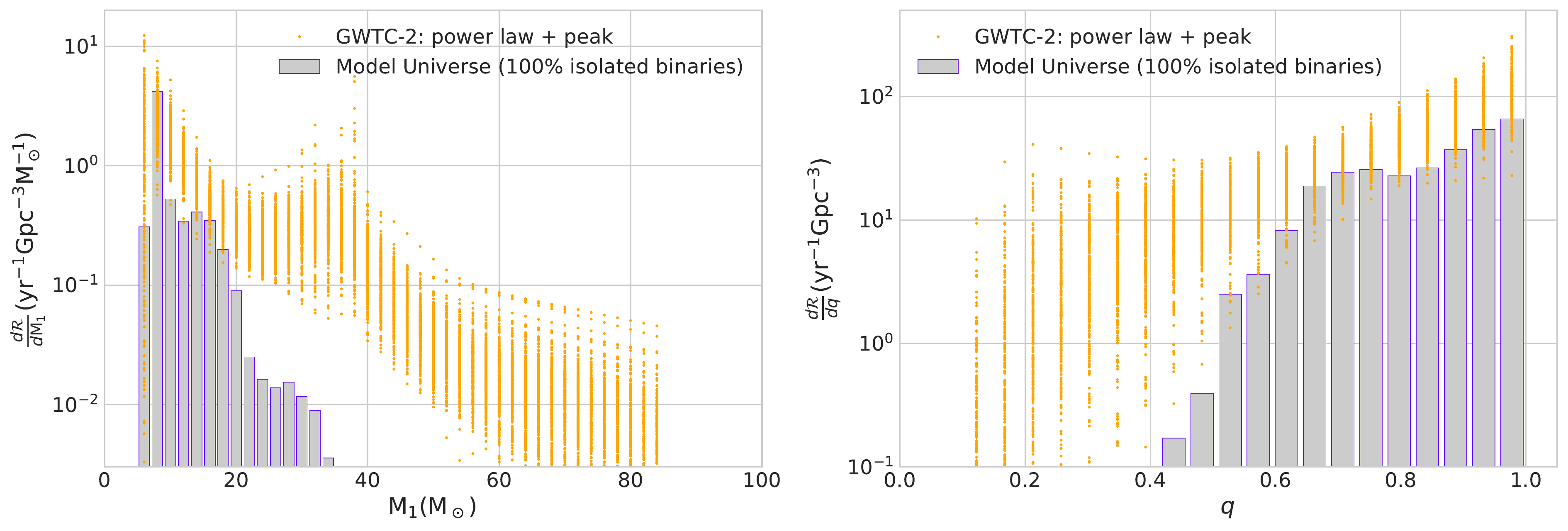}\\
\includegraphics[width=15.0cm,angle=0]{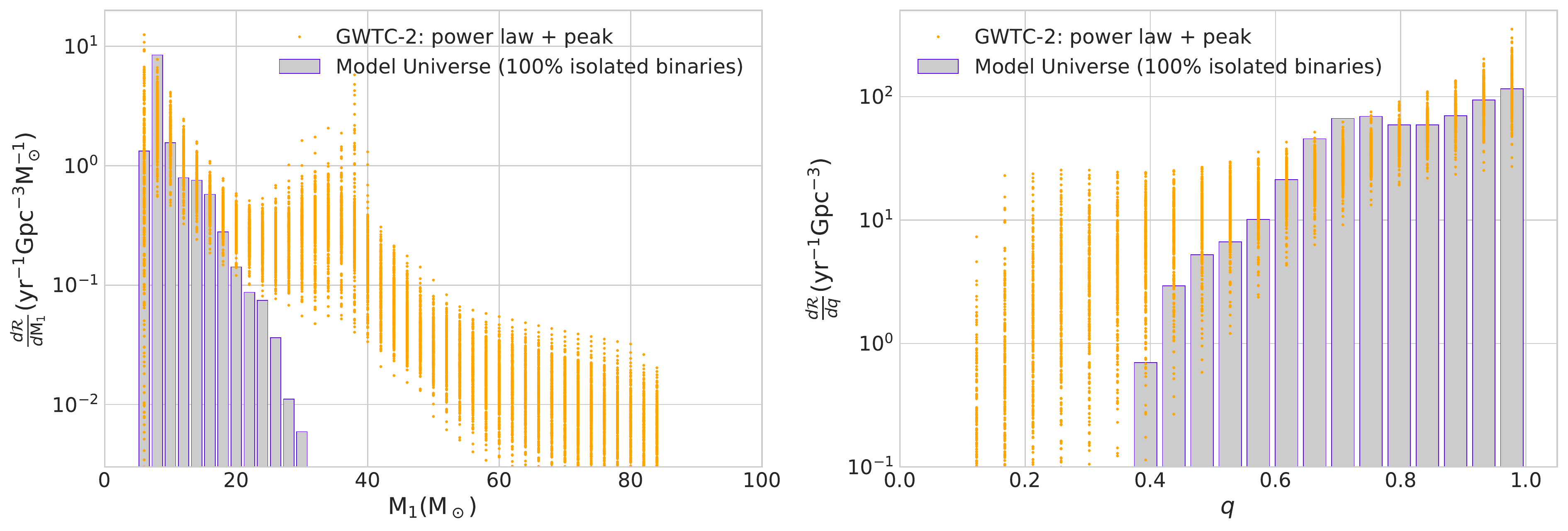}
	\caption{The filled histogram gives the present-day, differential intrinsic merger rate density (Y-axis) of BBHs,
	as obtained from Model Universe stellar populations (Sec.~\ref{popsynth}), as a function of merger primary mass
	(left panels) and mass ratio (right panels) along the X-axis. The orange dots are random draws (300 per
	bin) of the posteriors of BBH differential intrinsic merger rate densities as obtained from
	the LVK GWTC-2 \citep[][their power law + peak model]{Abbott_GWTC2_prop}.
	The top-row panels correspond to the hypothetical case where the entire star formation in the universe
	occurs in the form of YMCs of $\gtrsim10^4\Ms$. The other two rows correspond to the hypothetical cases
	where the entire star formation in the universe occurs in the form of isolated (\ie, never
	interacting dynamically with each other) field binaries with CE efficiency parameters
	$\ace=1$ (middle row) and $\ace=3$ (bottom row).}
\label{fig:diffrate_pure}
\end{figure*}

\begin{figure*}
\centering
\textbf{Pure channel: YMC/OC and IB ($\ace=1$ and $3$)}\par\medskip
\includegraphics[width=5.8cm,angle=0]{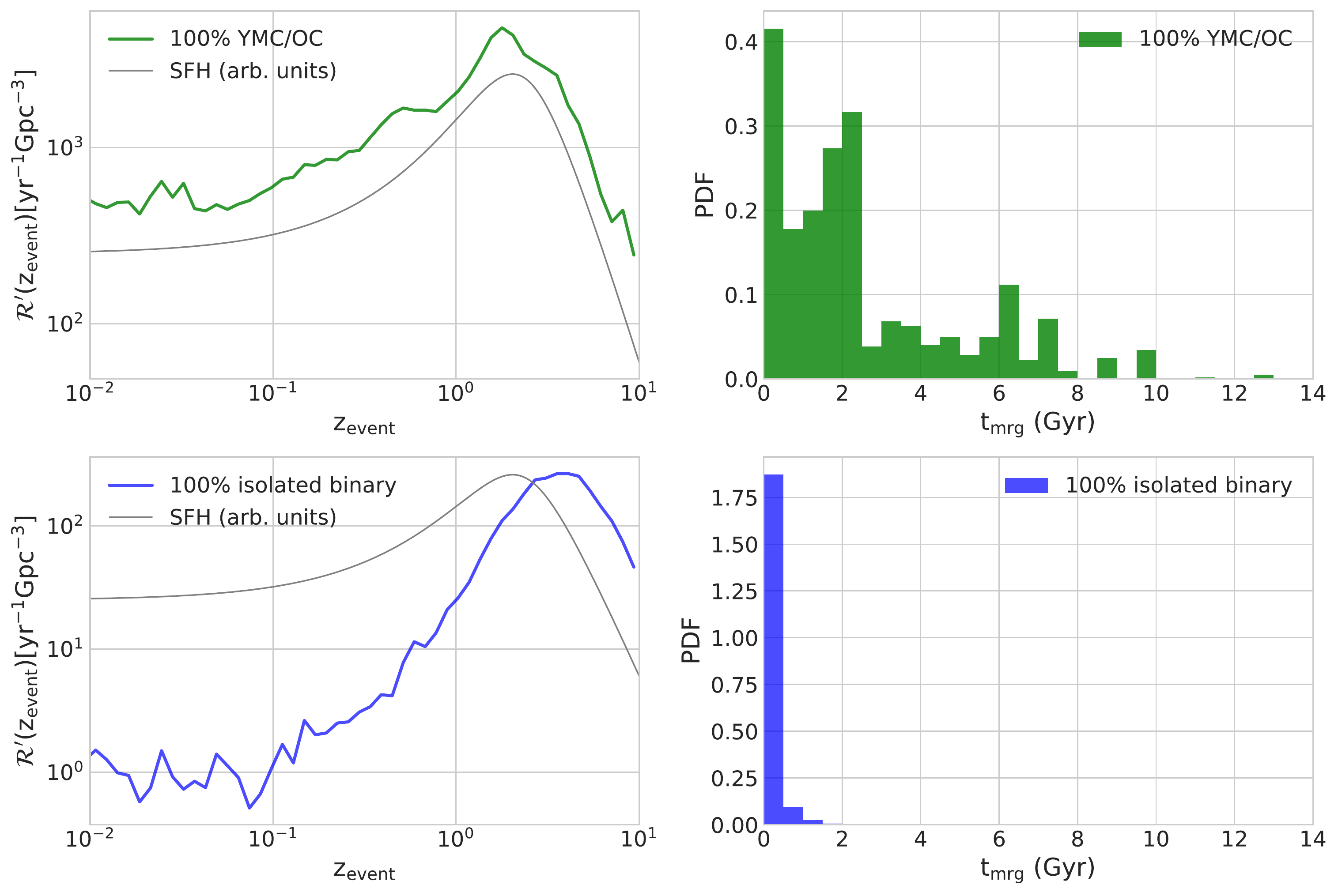}
\includegraphics[width=5.8cm,angle=0]{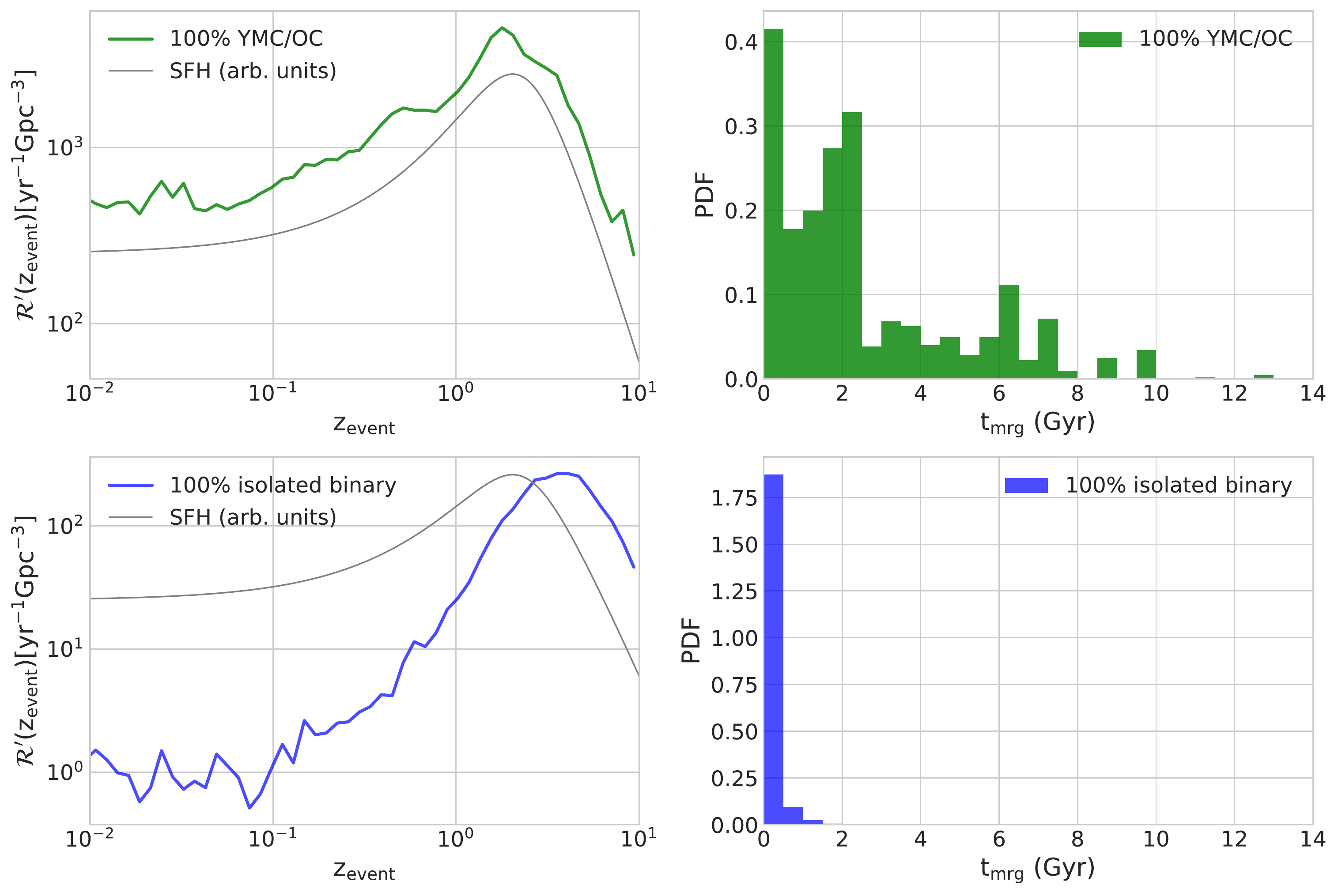}
\includegraphics[width=5.8cm,angle=0]{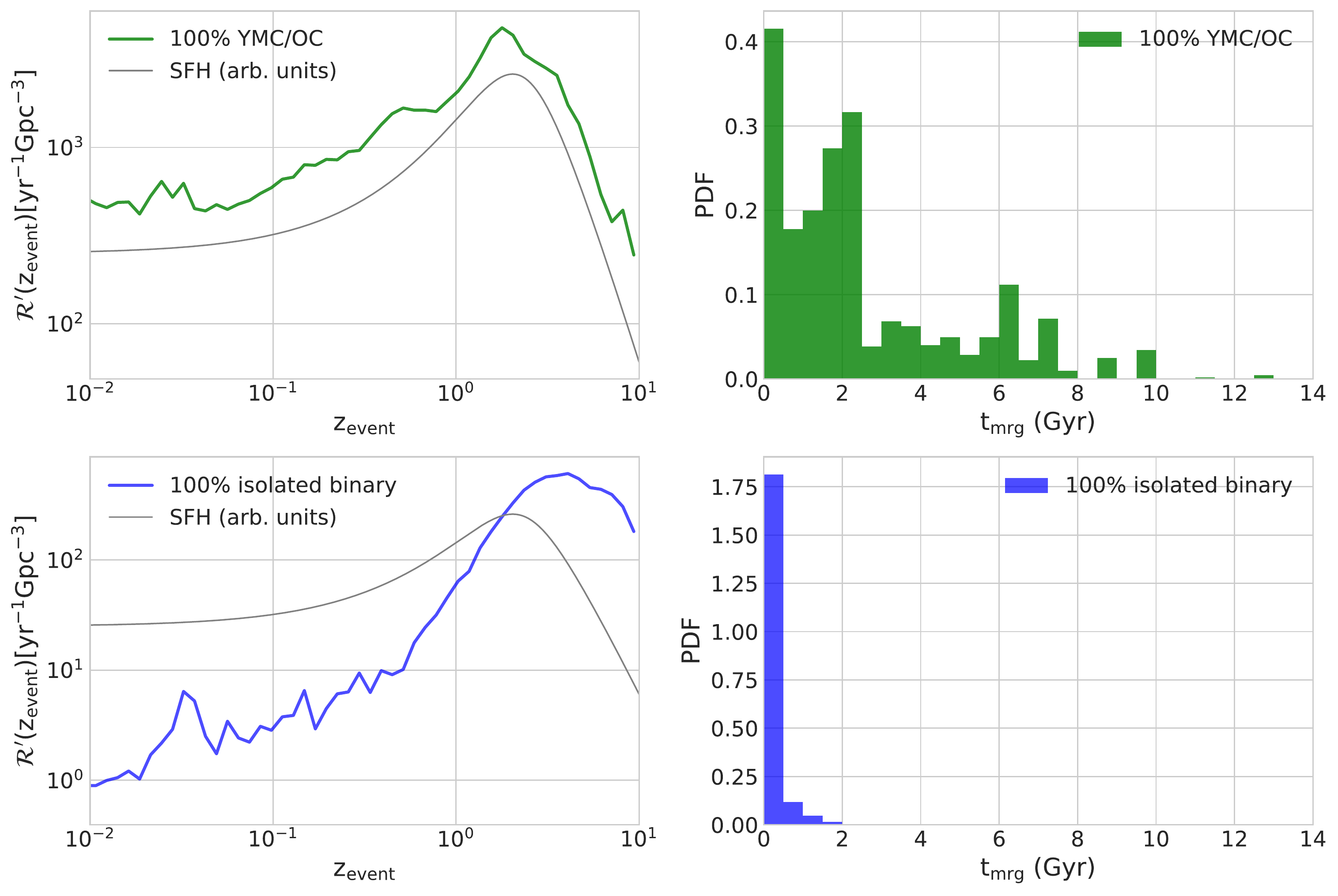}\\
\includegraphics[width=5.8cm,angle=0]{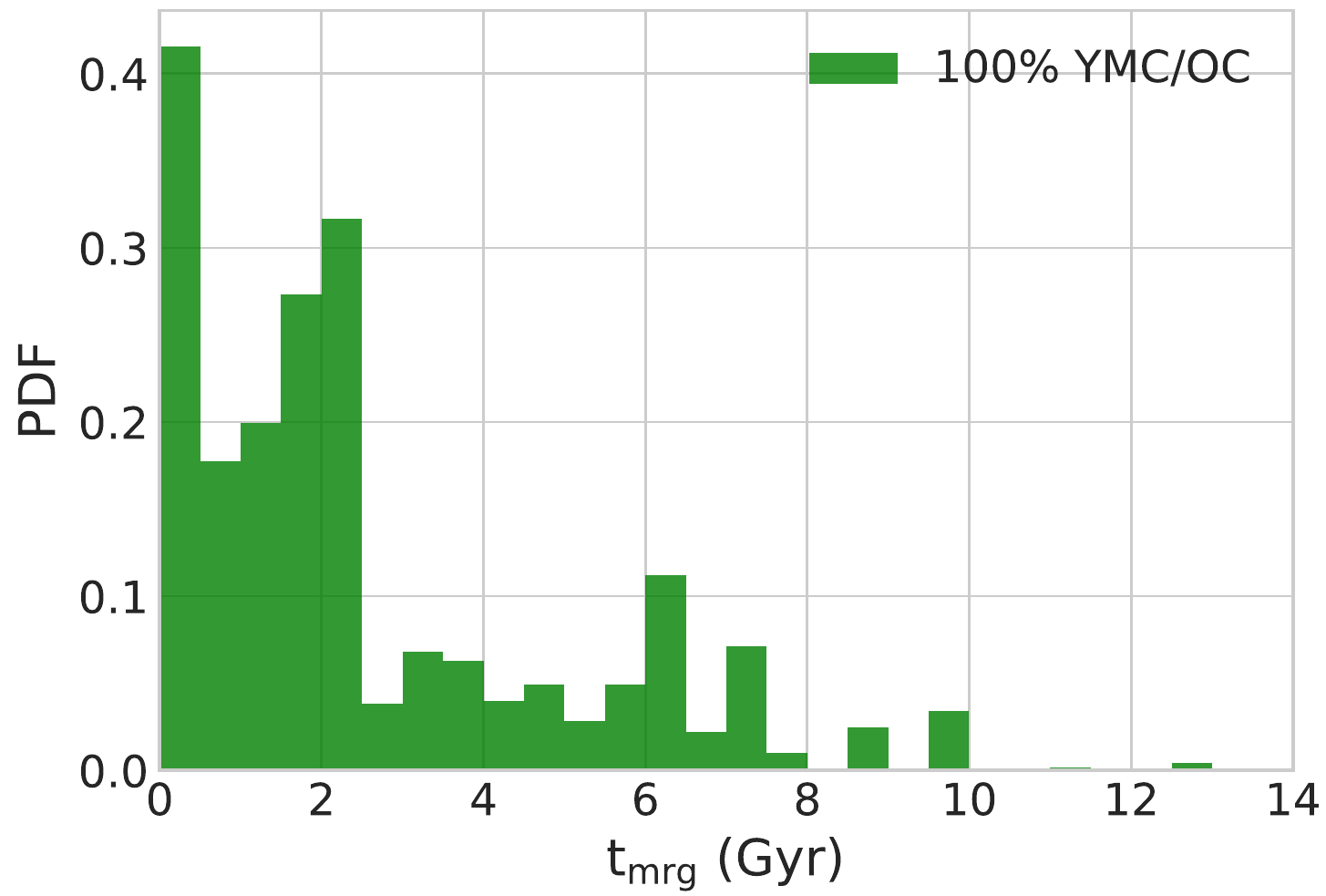}
\includegraphics[width=5.8cm,angle=0]{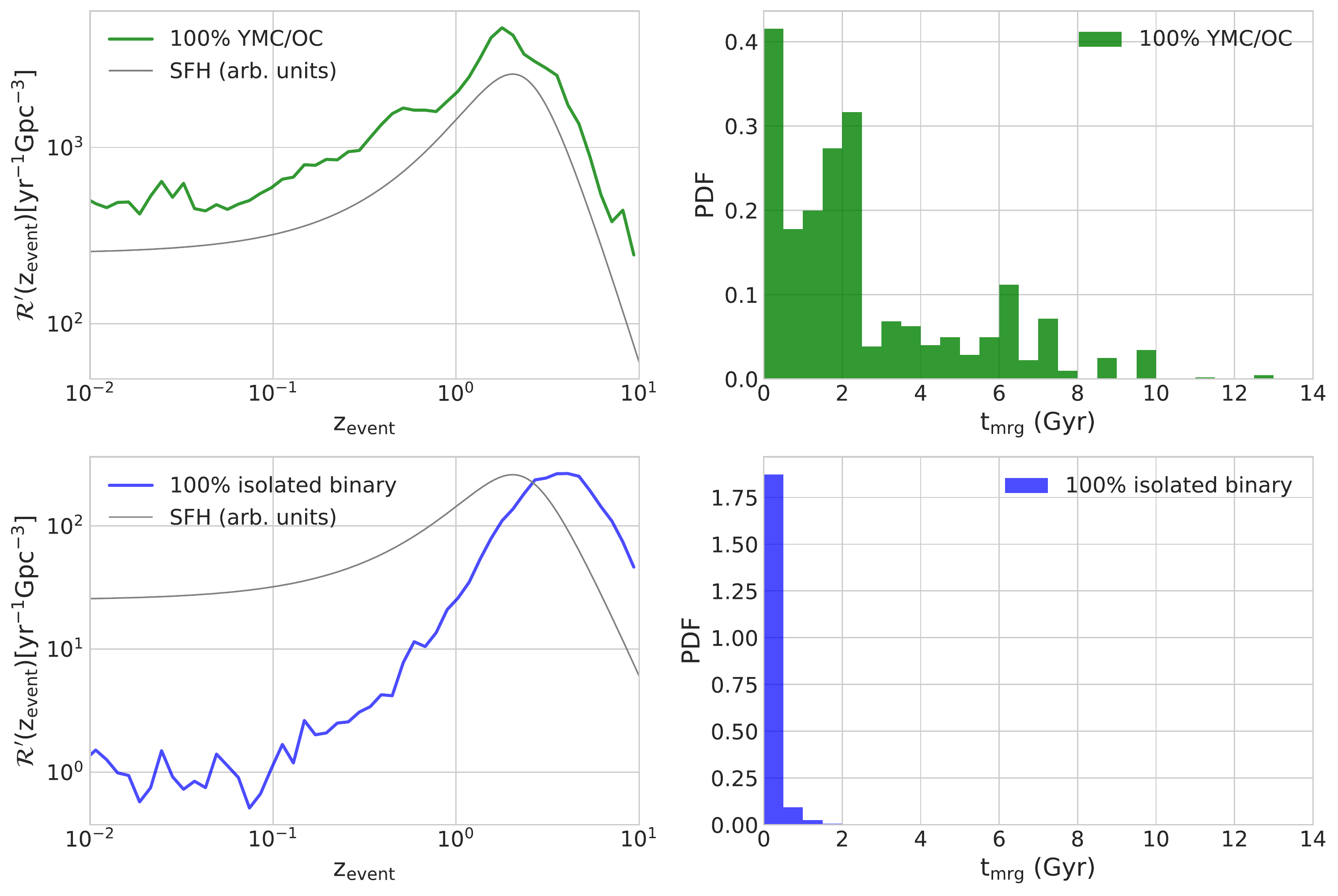}
\includegraphics[width=5.8cm,angle=0]{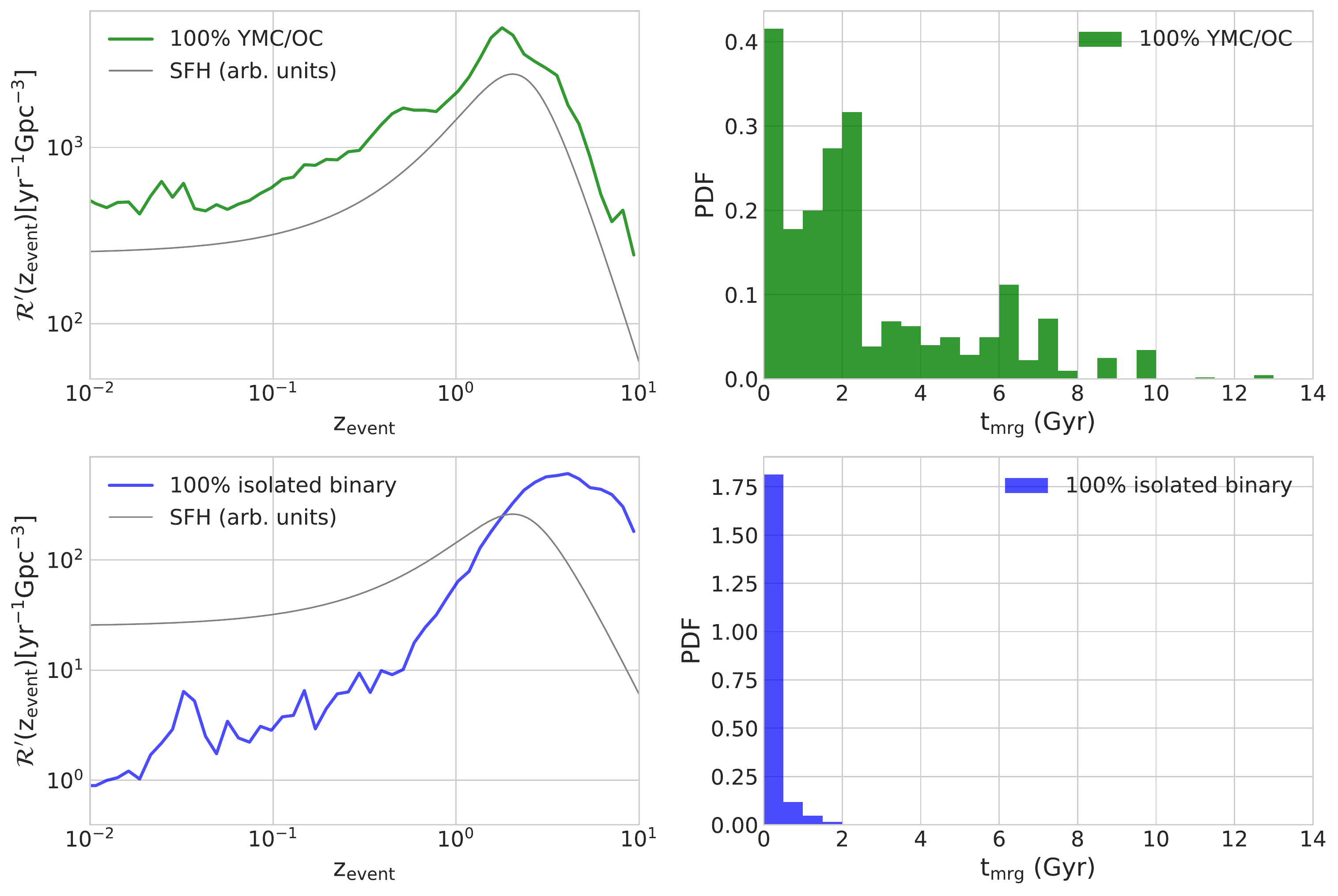}
	\caption{
	The cosmic evolution of BBH intrinsic merger rate density, $\rz(\zevnt)$ (top
	panels, green or blue line), and the distribution of merger delay times, $\tmrg$ (bottom panels),
	as obtained from Model Universe stellar populations (Sec.~\ref{popsynth}).
        The left-column panels correspond to the case where the entire star formation in the universe
	occurs in the form of YMCs of $\gtrsim10^4\Ms$. The other two columns correspond to the cases
	where the entire star formation in the universe occurs in the form of
	IBs with $\ace=1$ (middle column) and $\ace=3$ (right column).
	For visual comparison, the grey line (top panels) shows the variation of cosmic
	SFR with redshift (\cite{Madau_2017}; not to scale along the Y-axis).
	}
\label{fig:rz_pure}
\end{figure*}

To obtain the inherent dependence of Model Universe merger rate density
on redshift (the `cosmic merger rate density evolution'), $\rzi(\zevnt)$,
the merger events
from the sample population are binned according to their event redshifts,
$\zevnt$ (see above). The event count, $\Delta\nmrgi(\zevnt)$, over a redshift bin,
$\Delta\zevnt(\zevnt)$ around $\zevnt$, is then converted into the corresponding merger rate density,
$\rzi(\zevnt)$, by using Eqns.~\ref{eq:sfcrate}-\ref{eq:rate} and replacing
$2\delobs$ (Eqn.~\ref{eq:sfcrate}) by $\delage(\zevnt)$. Here, $\delage(\zevnt)$ is the
universe-age interval corresponding to the redshift interval $\Delta\zevnt(\zevnt)$.
Note that $\rzi(\zevnt)$ does not include light travel time (which is relevant only
for presently observed events)
\footnote{Hence, $\rzi(\zevnt)$ is independent of the $\zmax$ chosen in the
population synthesis exercise.}
but still incorporates merger delay time and
cosmic star-formation and metallicity evolutions.

Although for $\Delta\nmrgi(\zevnt)$ and $\rzi(\zevnt)$ star formation only up to
redshift $\zevnt$ is relevant, applying Eqns.~\ref{eq:sfcrate}-\ref{eq:rate}
is still valid since, for a sufficiently large $\nsampi$,
\beq
\frac{\msampi(\zevnt)}{\msampi} =
 \frac{\int_{t(z=10)}^{t(z=\zevnt)}\sfh(z(t))dt}{\int_{t(z=10)}^{t(z=0)}\sfh(z(t))dt}.
\label{eq:msamprop}
\eeq
Eqn.~\ref{eq:msamprop} assumes that the same (effective) fraction of star formation
goes into a specific stellar population type, I, throughout the cosmic history.
This assumption will be taken throughout this work. As in Ref.~\cite{Banerjee_2021},
100 equal-sized bins over $0\leq\zevnt\leq10$ are used to construct $\rz(\zevnt)$. 
To avoid processing an excessive volume of data over the large range of
$\zevnt$, $\nsampymc=2\times10^5$ and $\nsampiso=2\times2500$ are used for this
purpose.

In this work, redshift, comoving distance, and light travel time are interrelated
(based on a lookup table \cite{Wright_2006})
according to the $\Lambda$CDM cosmological framework \citep{Peebles_1993,Narlikar_2002}.
The cosmological constants from the latest Planck results
($H_0=67.4\kmps{\rm~Mpc}^{-1}$, $\Omega_{\rm~m}=0.315$, and flat Universe
for which $\thub=13.79{\rm~Gyr}$ \cite{Planck_2018}) are applied. Unless otherwise
stated (see Sec.~\ref{univ_2ch}), the `moderate-Z' \citep{Chruslinska_2019}
version of the cosmic metallicity  evolution is used.

\begin{table*}
\caption{Present-day, intrinsic, merger rate density and merger efficiency of BBHs,
	as obtained from the Model Universe, corresponding to the hypothetical
	cases where $100\%$ of the star formation takes place in the form of YMCs
        of $\gtrsim10^4\Ms$ or isolated (\ie, never
	interacting dynamically with each other) field binaries.
	The values
	in the second and third sections of this table are for `low-Z' (first row),
	`moderate-Z' (second row), and `high-Z' (third row)
	cosmic metallicity evolutions \citep{Chruslinska_2019} (Sec.~\ref{univ_2ch}).
	}
\label{tab_rates}
\centering
\begin{tabular}{l|cc}
	\hline
	Channel                     & Merger rate density $[\peryg]$  & Merger efficiency $[\Ms^{-1}]$ \\
	\hline
	100\% YMC/OC                &  $\rymc=1635.5$                  & $\etaymc=4.30\times10^{-5}$   \\
	100\% IB ($\ace=1$)         &  $\riso=13.1$                    & $\etaiso=1.36\times10^{-6}$   \\
	100\% IB ($\ace=3$)         &  $\riso=28.2$                    & $\etaiso=3.19\times10^{-6}$   \\
	\hline
	                            &                                  & $\etaymc=4.49\times10^{-5}$   \\
	100\% YMC/OC                &  $\langle\rymc\rangle=1618.1$    & $\etaymc=4.30\times10^{-5}$   \\
	                            &                                  & $\etaymc=4.23\times10^{-5}$   \\
	\hline
	                            &                                  & $\etaiso=3.83\times10^{-6}$   \\
	100\% IB ($\ace=3$)         &  $\langle\riso\rangle=31.6$      & $\etaiso=3.17\times10^{-6}$   \\
	                            &                                  & $\etaiso=2.71\times10^{-6}$   \\
	\hline
\end{tabular}
\end{table*}

Fig.~\ref{fig:diffrate_pure} shows the present-day differential merger rate densities
with respect to merger primary mass, $\mone$ (left panels), and merger mass ratio, $q$ (right panels),
for the hypothetical Model Universes with 100\% YMC/OC (top row) and
100\% IB of $\ace=1$ (middle row) and $\ace=3$ (bottom row). In a universe where
all of the star formation converts into pc-scale, gas-free YMCs of $\gtrsim10^4\Ms$,
$d\rymc/d\mone$ and $d\rymc/dq$ would greatly exceed than those
estimated from GWTC-2 \citep{Abbott_GWTC2_prop}, as Fig.~\ref{fig:diffrate_pure} suggests.
On the other hand, the IB counterpart of this universe would produce
GWTC-2-like merger rates but the corresponding $d\riso/d\mone$ would sharply fall below 
the GWTC-2 differential rates for $\mone\gtrsim20\Ms$. The resulting total BBH merger rates
corresponding to the two universes are quoted in Table~\ref{tab_rates}.
Table~\ref{tab_rates} also quotes the merger efficiencies, $\etaymc$ and $\etaiso$,
of the two universes. In this work, merger efficiency is simply defined as the
number of mergers per unit mass of star formation in a given universe,
averaged over redshift and metallicity (\ie, it refers to the universe as a whole
rather than a specific type of cluster or a binary population).

Fig.~\ref{fig:rz_pure} shows the cosmic merger rate density evolutions (left
panels) and merger delay time distributions (right panels) for the universes with 100\% YMC/OC (top row) and
100\% IB of $\ace=1$ (middle row) and $\ace=3$ (bottom row). This figure clearly
illustrates the stark difference between the delay times, $\tmrg$s, of the BBH mergers
originating from YMC/OCs (dynamically-assembled mergers) and IBs (binary-evolutionary mergers).
The $\tmrg$s from the IBs are mostly concentrated within 500 Myr with a tail in their
distribution extending up to 2 Gyr. The predominance of short delay times,
in combination with higher formation efficiency of tight BBHs (those with $\tmrg<\thub$)   
at lower metallicities \citep{Giacobbo_2018,Baibhav_2019} that are more dominant at higher $z$,
translates into $\rzib(\zevnt)$ peaking at an epoch earlier than the cosmic-SFH peak.
This result has also been found in other  
recent works that apply similar binary population synthesis approaches
(\eg, \citep{Baibhav_2019}).
The exact form of $\rzib(\zevnt)$ depends, therefore, on
the adopted cosmic metallicity evolution: those in Fig.~\ref{fig:rz_pure} 
corresponds to that in Ref.~\cite{Chruslinska_2019} (their `moderate-Z' dependence)
as incorporated here.
Note that the overall nature of $\rzib(\zevnt)$ for $\ace=1$ and 3, as obtained here,
are similar to those obtained by
other recent, similar binary population synthesis studies
(\eg, \citep{Baibhav_2019,Santoliquido_2020}).
In contrast, the majority of the $\tmrg$s from the YMC/OCs are of $\lesssim2$ Gyr
with a tail in their distribution reaching up to $\thub$. The longer $\tmrg$s
result in $\rzymc(\zevnt)$ maximizing at a more recent epoch, matching with the SFH peak
(see Ref.~\cite{Banerjee_2021} for further discussions).

\section{Merger rate density of stellar-mass binary black holes from young massive clusters,
open clusters, and isolated binaries}\label{res}

\begin{figure*}
\centering
\textbf{2-channel}\par\medskip
\includegraphics[width=8.0cm,angle=0]{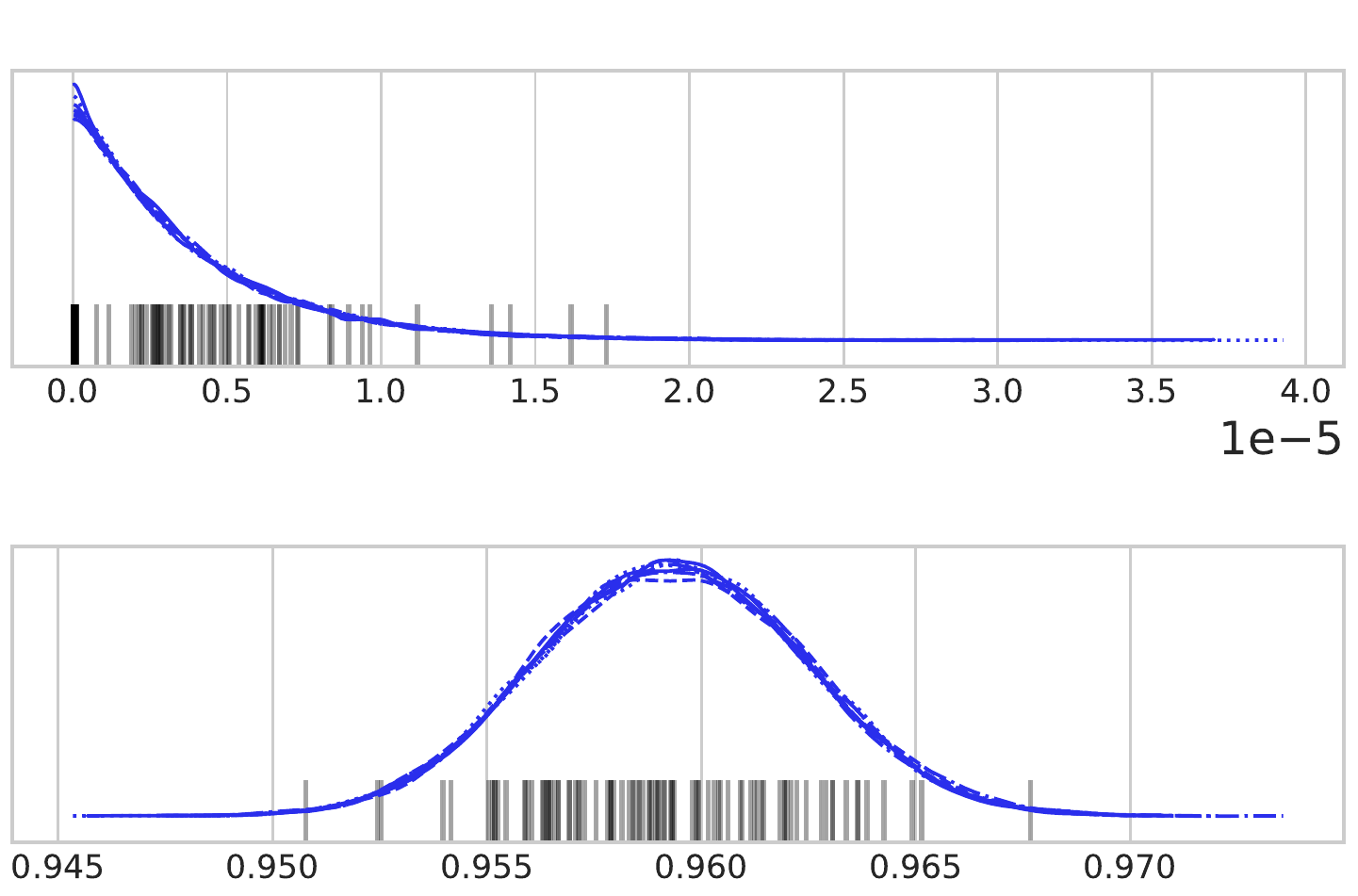}
\includegraphics[width=8.0cm,angle=0]{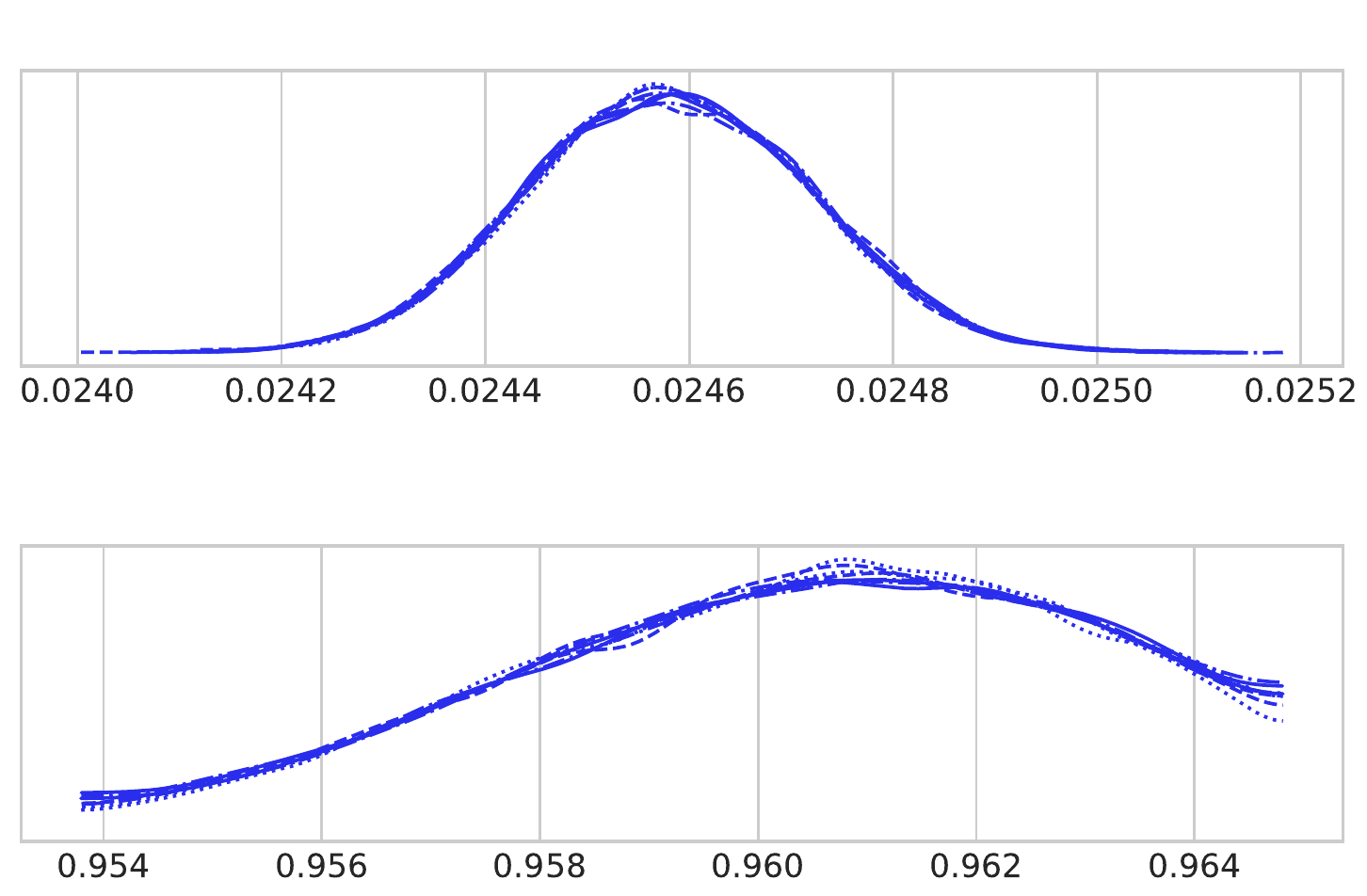}
	\caption{Examples of posterior distributions of $\fymc$ (top panels) and $\fobin$
	(bottom panels) in the first and second iterations (left and right panels,
	respectively) of the Bayesian regression analysis described in Sec.~\ref{res}. This
	demonstration corresponds to the use of the third moments of $d\rate/d\mone$
	and $d\rate/dq$ and IB evolution with $\ace=3$.
        In all panels,
	the posterior distributions from the 8 MCMC chains (Sec.~\ref{res}) are plotted separately,
	demonstrating good convergence.
	}
\label{fig:post_2ch}
\end{figure*}

\begin{figure*}
\centering
\textbf{1-channel}\par\medskip
\includegraphics[width=8.0cm,angle=0]{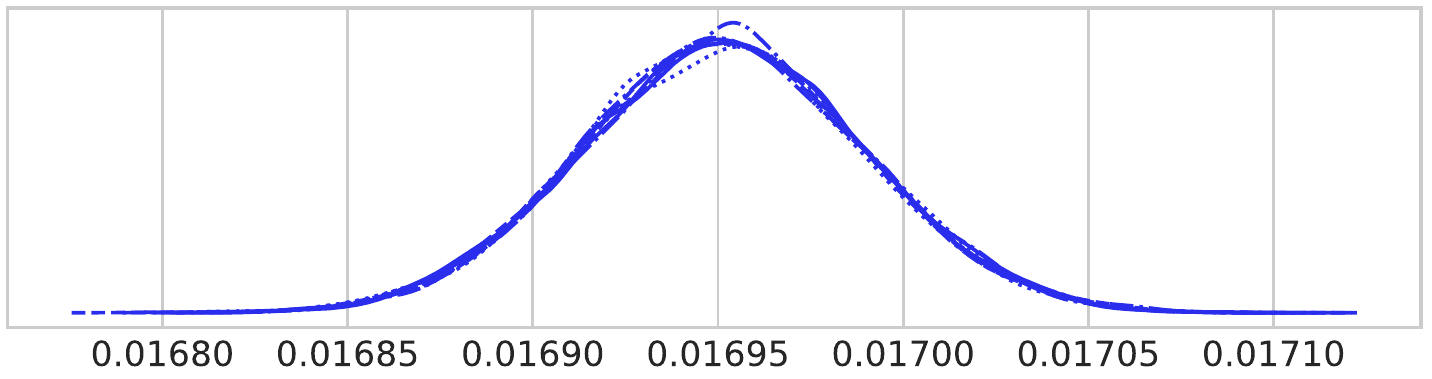}
\includegraphics[width=8.0cm,angle=0]{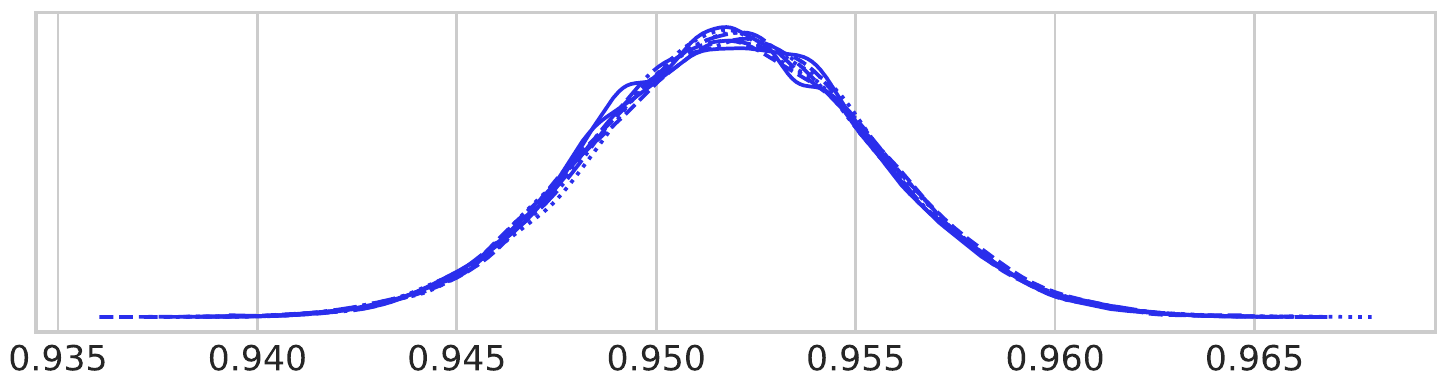}
	\caption{{\bf Left:} posterior distribution of YMC formation efficiency,
	$\fymc$, in the Model Universe assuming that the observed, present-day BBH merger rate density
	\citep{Abbott_GWTC2_prop} is due to the only channel of dynamical interactions inside YMCs/OCs
	of the universe.
	{\bf Right:} posterior distribution of binary fraction among OB stars,
	$\fobin$, in the Model Universe assuming that the observed, present-day BBH merger rate density
	is due to the only channel of isolated evolution of massive binaries
	of the universe. In both panels,
	the posterior distributions from the 8 MCMC chains (Sec.~\ref{res}) are plotted separately,
	demonstrating good convergence. These posterior distributions correspond
	to uninformed priors of $\fymc$ and $\fobin$.}
\label{fig:post_1ch}
\end{figure*}

\begin{table*}
\caption{The mean value of the posteriors of YMC formation efficiency, $\langle\fymc\rangle$,
	and that of the OB-star binary fraction, $\langle\fobin\rangle$, for the Model Universe
	in the various cases indicated in the left column. The posteriors of $\fymc$ and $\fobin$
	are obtained from their uninformed priors by applying a two-stage Bayesian regression
	method as described in Sec.~\ref{res}. $\ace$ is the CE efficiency parameter applied in
	the isolated-binary (IB) evolution and $p$ is the order of the moment of the
	present-day differential merger rate density used in the Bayesian analysis. The values
	in the final section of this table correspond to considering the `low-Z', `moderate-Z', and `high-Z'
	cosmic metallicity evolutions \citep{Chruslinska_2019} together (with equal weights).}
\label{tab_vals}
\centering
\begin{tabular}{l|cc}
	\hline
	Channel                      &     $\langle\fymc\rangle$    &  $\langle\fobin\rangle$  \\
	\hline
	YMC/OC ($p=0$)               &       $1.69\times10^{-2}$    &      $0.0$               \\
	\hline
	IB ($\ace=1, p=0$)           &           $0.0$              &  $1.00$                  \\ 
	IB ($\ace=3, p=0$)           &           $0.0$              &  $9.52\times10^{-1}$     \\
	\hline
 	YMC/OC + IB ($\ace=1, p=0$)  & $7.21\times10^{-3}$          &  $1.00$                  \\
	YMC/OC + IB ($\ace=1, p=1$)  & $1.04\times10^{-2}$          &  $1.00$                  \\
	YMC/OC + IB ($\ace=1, p=2$)  & $1.63\times10^{-2}$          &  $1.00$                  \\
	YMC/OC + IB ($\ace=1, p=3$)  & $2.47\times10^{-2}$          &  $1.00$                  \\
	YMC/OC + IB ($\ace=1, p=4$)  & $3.74\times10^{-2}$          &  $1.00$                  \\
	\hline
 	YMC/OC + IB ($\ace=3, p=0$)  & $1.51\times10^{-5}$          &  $9.42\times10^{-1}$     \\
	YMC/OC + IB ($\ace=3, p=1$)  & $6.34\times10^{-3}$          &  $9.44\times10^{-1}$     \\
	YMC/OC + IB ($\ace=3, p=2$)  & $1.50\times10^{-2}$          &  $9.12\times10^{-1}$     \\
	YMC/OC + IB ($\ace=3, p=3$)  & $2.46\times10^{-2}$          &  $9.60\times10^{-1}$     \\
	YMC/OC + IB ($\ace=3, p=4$)  & $3.66\times10^{-2}$          &  $9.62\times10^{-1}$     \\
	\hline
	YMC/OC + IB ($\ace=3, p=3$)  & $1.48\times10^{-2}$          &  $8.69\times10^{-1}$     \\
	YMC/OC + IB ($\ace=3, p=4$)  & $2.23\times10^{-2}$          &  $9.06\times10^{-1}$     \\
	\hline
\end{tabular}
\end{table*}

Having obtained, as in Sec.~\ref{popsynth}, the present-day BBH merger rate densities
and their cosmic evolutions for the hypothetical YMC-only and IB-only universes,
they can be scaled and combined with respect to astrophysical quantities
to obtain the merger rate (evolution) in a more realistic universe. The
differential and total rates from YMC/OCs are proportional to the
YMC formation (as fully-assembled, gas-free YMCs) efficiency, $\fymc$
(the YMC-only universe corresponds to $\fymc=1$). The rates from
IBs are proportional to OB-star binary fraction $\fobin$
(the IB-only universe corresponds to $\fobin=1$). If both formation
channels contribute to the universe's BBH mergers, then combined
rates are given by
\beq
\frac{d\rate}{dX}(X) = \fymc\frac{d\rymc}{dX}(X) + \fobin\frac{d\riso}{dX}(X)
\label{eq:rates_2ch}
\eeq
and
\beq
\rz(\zevnt) = \fymc\rzymc(\zevnt) + \fobin\rzib(\zevnt).
\label{eq:rz_2ch}
\eeq
This simple linear combination, of course, assumes that $\fymc$ and $\fobin$
can be represented with constant effective values throughout the cosmic history.

In this study, $\fymc$ and $\fobin$ are determined through a Bayesian-regression
approach. The results in Sec.~\ref{popsynth} (see Figs.~\ref{fig:diffrate_pure} and \ref{fig:rz_pure})
suggest that it is important to incorporate the detailed form of the
differential rate distributions in determining the relative contributions
of various merger channels. Therefore, the likelihood functions are constructed
based on various moments of the differential rate density functions 
from the Model Universe and GWTC-2. The $p$-th moment of the differential merger rate density
function defined over an interval $X\in[X1,X2]$ is
\beq
\mupi\equiv\int_{X1}^{X2}X^p\frac{d\ratei}{dX}(X)dX.
\label{eq:momdef}
\eeq
Therefore, the moment of the combined distribution is (using Eqn.~\ref{eq:rates_2ch})
\begin{equation}
\mup=\int_{X1}^{X2}X^p\frac{d\rate}{dX}(X)dX
    = \fymc\mupymc + \fobin\mupiso.
\label{eq:mu_2ch}
\end{equation}

In the present Bayesian approach, $\fymc\in[0,1]$ and $\fobin\in[0,1]$ are taken to be free
parameters to be estimated based on merger rate densities from GWTC-2
and the Model Universe. The elements of the likelihood function are taken to
be of the normal form and the priors of $\fymc$ and $\fobin$ are taken to be unbiased. Hence,
Bayes theorem \citep{Grinstead_2012} becomes
\begin{equation}
\begin{aligned}
	& P(\fymc,\fobin|\muobs) =  \\
	&  \frac{\lhood(\fymc,\fobin)P(\fymc)P(\fobin)}
	{\int\limits_{\fymc}\int\limits_{\fobin}\lhood(\fymc,\fobin)P(\fymc)P(\fobin)d\fymc d\fobin}.
\end{aligned}
\label{eq:bayes}
\end{equation}
Here, $P(\fymc,\fobin|\muobs)$ is the (joint) posterior probability distribution of
$\fymc$ and $\fobin$. $P(\fymc)$ and $P(\fobin)$ are the prior probability
distributions of $\fymc$ and $\fobin$, both of which are taken to be
uniform over $[0,1]$, $\unif(0,1)$, at the initial iteration (see below).
$\lhood(\fymc,\fobin)$ is the likelihood function given by
\begin{equation}
\begin{aligned}
	\lhood(\fymc,\fobin)  =  P(\muobs|\fymc,\fobin) & \\
	 =  \prod_i^{\nobs}\norm\left[\mup(\fymc,\fobin)-\muobsi,\Delta\mup(\fymc,\fobin)\right]. & 
\end{aligned}
\label{eq:lhood}
\end{equation}

Here, $\norm[\mu,\sigma]$ represents a normal probability distribution with
mean $\mu$ and variance $\sigma^2$. $\muobsi\in\muobs$ are the moments
of the GWTC-2 intrinsic differential merger rate densities.
To obtain these, $\nobs=300$ random values of the posteriors of GWTC-2
differential merger rate densities (their `power law + peak' model)
\footnote{The GWTC-2 data utilized in this work is publicly available at
the URL \url{https://dcc.ligo.org/LIGO-P2000434/public}.}
are chosen at each bin (the orange dots in the
panels of Fig.~\ref{fig:diffrate_pure}).
The resulting $\nobs$ different distributions then give
$\nobs$ different values, $\muobsi$. $\mup(\fymc,\fobin)$ is the combined
Model Universe moment as given by Eqn.~\ref{eq:mu_2ch}. $\Delta\mup(\fymc,\fobin)$
is a measure of the variance of the Model Universe moment
given by (follows from Eqn.~\ref{eq:mu_2ch})
\beq
\Delta\mup = \fymc\Delta\mupymc + \fobin\Delta\mupiso.
\label{eq:delmu_2ch}
\eeq
(For brevity, $\mup(\fymc,\fobin)$ and $\Delta\mup(\fymc,\fobin)$ will hereafter be 
written without the arguments.)
$\Delta\mupi$ comprises errors from all the
bins, \ie (following from Eqn.~\ref{eq:momdef}; taking
idealized parameter estimation in the Model Universe implying $\Delta X=0$),
\beq
\Delta\mupi=\int_{X1}^{X2}X^p\left[\Delta\frac{d\ratei}{dX}(X)\right]dX.
\label{eq:delmupi}
\eeq
In practice, $\Delta(d\ratei/dX)$ at a particular bin is determined
by stacking the outcomes of the independent sample-population trials
and taking the difference of the resulting maximum and minimum rates,
for that bin (Eqns.~\ref{eq:diffrate}-\ref{eq:ratepdf}).
For single
SFH, age-redshift, and metallicity-redshift dependencies,
as used in the Model Universe (Sec.~\ref{popsynth}),  
$\Delta(d\ratei/dX)$ is comparable to that due to the Poisson
error in the bin. However, larger variations would result
by incorporating astrophysical variations, as demonstrated
below (Sec.~\ref{univ_2ch}).

To take into account the present-day differential merger rate density distributions
with respect to both $q$ and $\mone$, a two-step
procedure is followed. First, the posteriors of $\fymc$ and $\fobin$
are obtained (Eqn.~\ref{eq:mu_2ch}-\ref{eq:delmupi}),
assuming their $\unif(0,1)$ prior distributions,
by considering the moments of only the rate distributions with respect to $q$.
The resulting posteriors of $\fobin$ are then treated as priors
of the same in the next iteration. In this following iteration,
the posteriors of $\fymc$ and $\fobin$ are redetermined by considering
the moments of only the rate distributions with respect to $\mone$
and resetting the prior distribution of $\fymc$ to $\unif(0,1)$.
The resulting posteriors are taken to be final.

The construction of the likelihood function and the sampling
of the posteriors are done utilizing the {\tt Python} package
$\pymc$ \citep{Martin_2018}, using its {\tt Uniform}, {\tt Normal}, {\tt Interpolated},
and {\tt sample} utilities. The posteriors are obtained by applying a
(Hamiltonian) Markov Chain Monte Carlo (hereafter MCMC) approach that 
employs the {\tt No U-turn Sampler} of the package. $8\times12500$ posterior samples
(plus $8\times1500$ tuning iterations) are drawn from 8 independent MCMC chains.
For each chain, the first 1000 values are discarded (or `burned') 
to avoid incorporating spurious values in the posterior distributions.

This two-stage procedure is inspired by the fact that despite
the large difference in amplitude,
$d\rymc/dq$ and $d\riso/dq$ are of similar shape and truncations
unlike $d\rymc/d\mone$ and $d\riso/\mone$ which are largely
dissimilar (Sec.~\ref{popsynth}; Fig.~\ref{fig:diffrate_pure}).
Hence, one can first `learn' about $\fobin$ from the $q$ distributions
and then further refine the inferences on $\fymc$ and $\fobin$
from the $\mone$ distributions. That way, zero
to a few divergences among the MCMC chains (as reported by the sampler
summary) are always obtained at the end of the second sampling iteration. 
The posterior distributions of $\fymc$ and $\fobin$ from the two
iterations are shown in the example of Fig.~\ref{fig:post_2ch},
which shows excellent agreement between the final distributions obtained
from the 8 MCMC chains separately. Such MCMC traces are plotted using the {\tt ArviZ}
package \citep{Martin_2018}. The rest of the figures in
this paper are plotted using {\tt Matplotlib}\citep{Hunter_2007}
\footnote{\url{https://matplotlib.org}}.

A simpler, single-iteration version of the above procedure is applied
for a one-channel universe, \ie, where BBH mergers are produced from either
YMC/OC or field binaries having fractions $\fymc$ or $\fobin$, respectively.  
In other words, Eqns.~\ref{eq:mu_2ch} and \ref{eq:delmu_2ch} have $\fobin=0$ or
$\fymc=0$, respectively.
In this case, only the zeroth moment (\ie, total rate) of the
$q$ distributions are utilized for a one-stage
estimation of $\fymc$ ($\fobin$) posteriors,
taking the prior distribution of $\fymc$ ($\fobin$) to be $\unif(0,1)$.
This exercise, typically, also yields good convergence.
Fig.~\ref{fig:post_1ch} shows such an example of posteriors from the
8 MCMC sampling chains.

Note that the present method is still preliminary and proof-of-concept.
In particular, no `hyper-parameter' is applied. Such parameters can be,
\eg, SFH slope, metallicity-redshift slope, cluster-structural parameters,
binary-physics parameters, BH-spins, that determine the probabilities of present-day merger
and detection beyond a signal-to-noise-ratio threshold.
In a future work, such a more complete Bayesian analysis and inference
\citep{Mandel_2019,Bouffanais_2021} will be explored. The present exercise,
although explicitly involves Bayes theorem and data from theoretical models
and from analyses of observed event parameters (specifically, GWTC-2
`power law + peak' intrinsic merger rate densities),
can be described as a `Bayesian regression' procedure.

\subsection{One-channel universe}\label{univ_1ch}

\begin{figure*}
\centering
\textbf{1-channel: YMC/OC}\par\medskip
\includegraphics[width=15.0cm,angle=0]{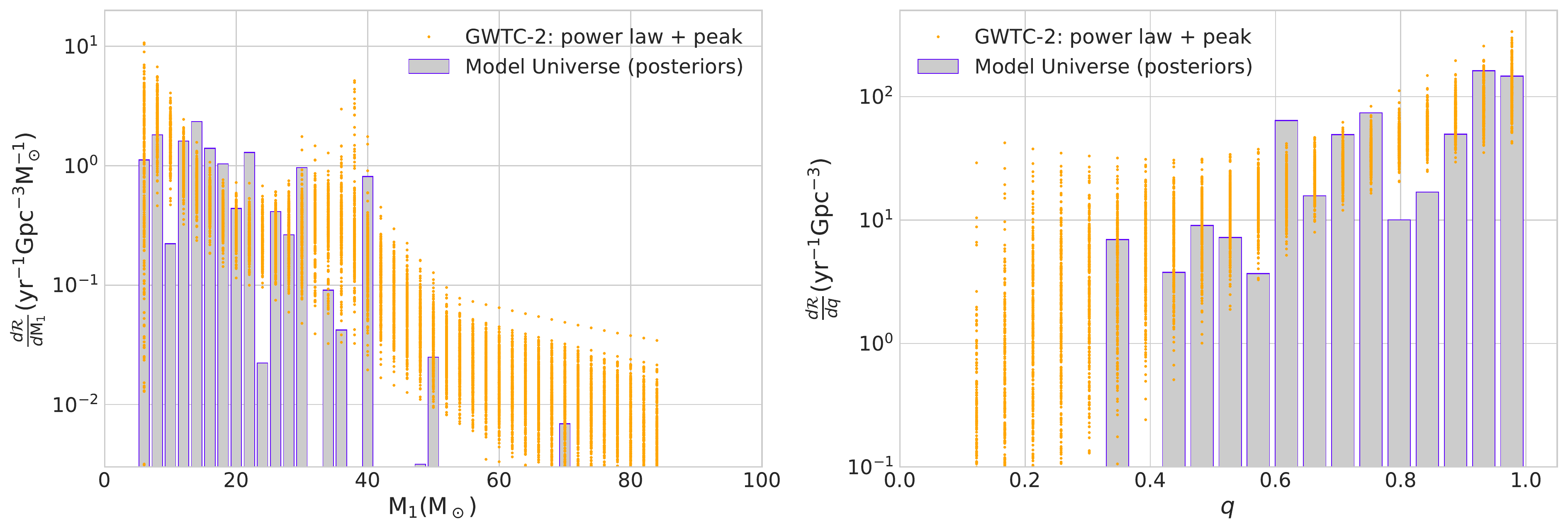}\\
\includegraphics[width=8.0cm,angle=0]{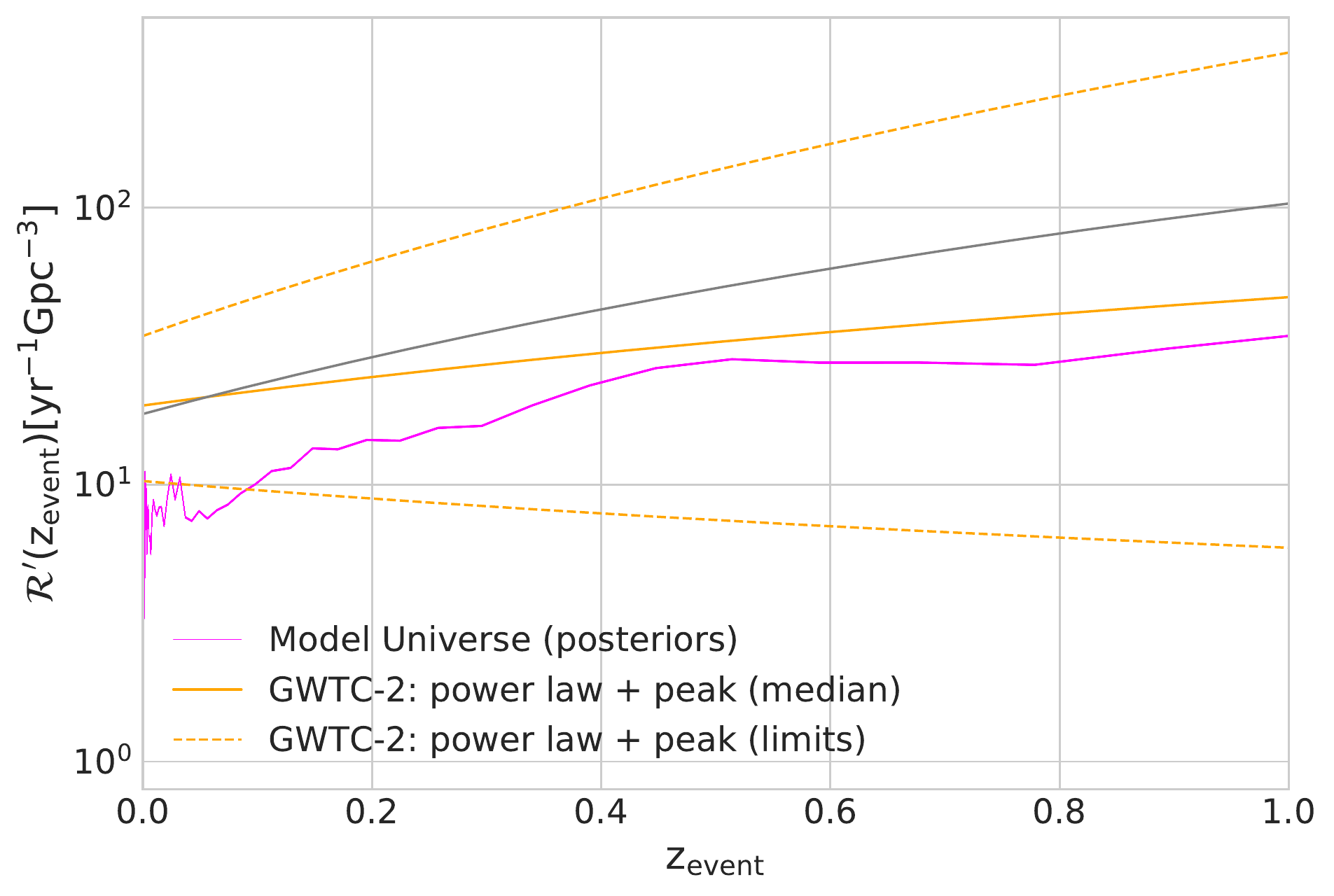}
	\caption{
	The Model Universe present-day, differential intrinsic merger rate density of BBHs (top panels;
	the legends are the same as in Fig.~\ref{fig:diffrate_pure}) and the cosmic evolution of BBH
	intrinsic merger rate density (bottom panel, magenta line), 
	assuming that the observed,
	present-day BBH merger rate density \citep{Abbott_GWTC2_prop} is due to the only channel of dynamical
	interactions inside YMCs/OCs of the universe. The Model Universe rates shown in these panels  
	correspond to 200 choices of the posteriors of $\fymc$ in the YMC-only universe, as shown
	in Fig.~\ref{fig:post_1ch}. The orange solid (dashed) line(s) in the bottom
	panel depict the median (90\%-credible limits) of the GWTC-2 cosmic merger
	rate density evolution: these lines are given by
	$\rz(\zevnt)=19.3_{-9.0}^{+15.1}\peryg(1+\zevnt)^{1.3_{-2.1}^{+2.1}}$
	\citep[][their power law + peak model]{Abbott_GWTC2_prop}.
	As in Fig.~\ref{fig:rz_pure}, the grey line (bottom panel) depicts the variation of cosmic
	SFR with redshift (arbitrary unit along the Y-axis).
	}
\label{fig:rates_1ch_ymc}
\end{figure*}

Fig.~\ref{fig:rates_1ch_ymc} (top panels) shows the $d\rate/d\mone$
and $d\rate/dq$
for the Model Universe with $\fobin=0$, \ie, when only YMC/OCs of the universe
produce BBH mergers. The Model Universe rates are plotted for 200 random choices
of the posteriors of $\fymc$. The mean of the $\fymc$ posteriors is stated
in Table~\ref{tab_vals}. The corresponding $\rz(\zevnt)$ (up to $\zevnt=1$)
is shown in the bottom panel of Fig.~\ref{fig:rates_1ch_ymc}. See the figure's
caption for further detail. Fig.~\ref{fig:rates_1ch_ymc} essentially
reproduces the results obtained in Ref.~\cite{Banerjee_2021}
but with differently-obtained normalization. It suggests that,
in principle, dynamical BBH mergers in moderate-mass YMCs and OCs in the Universe
alone can self-consistently explain the present-day, differential intrinsic BBH merger
rate density and the cosmic evolution of intrinsic BBH merger density,
as inferred from GWTC-2. However, for $\zevnt\lesssim0.2$, $\rz$ falls below
the GWTC-2 median by a few factors and reaches the GWTC-2 lower limit at
$\zevnt\approx0$ (Fig.~\ref{fig:rates_1ch_ymc}, lower panel).
Of course, the very high $\rymc$ 
of the 100\%-YMC universe (Sec.~\ref{popsynth}, Table~\ref{tab_rates})
results in the inference of the small mean $\langle\fymc\rangle\sim 10^{-2}$ (Table~\ref{tab_vals}).

\begin{figure*}
\centering
\textbf{1-channel: IB ($\ace=1$)}\par\medskip
\includegraphics[width=15.0cm,angle=0]{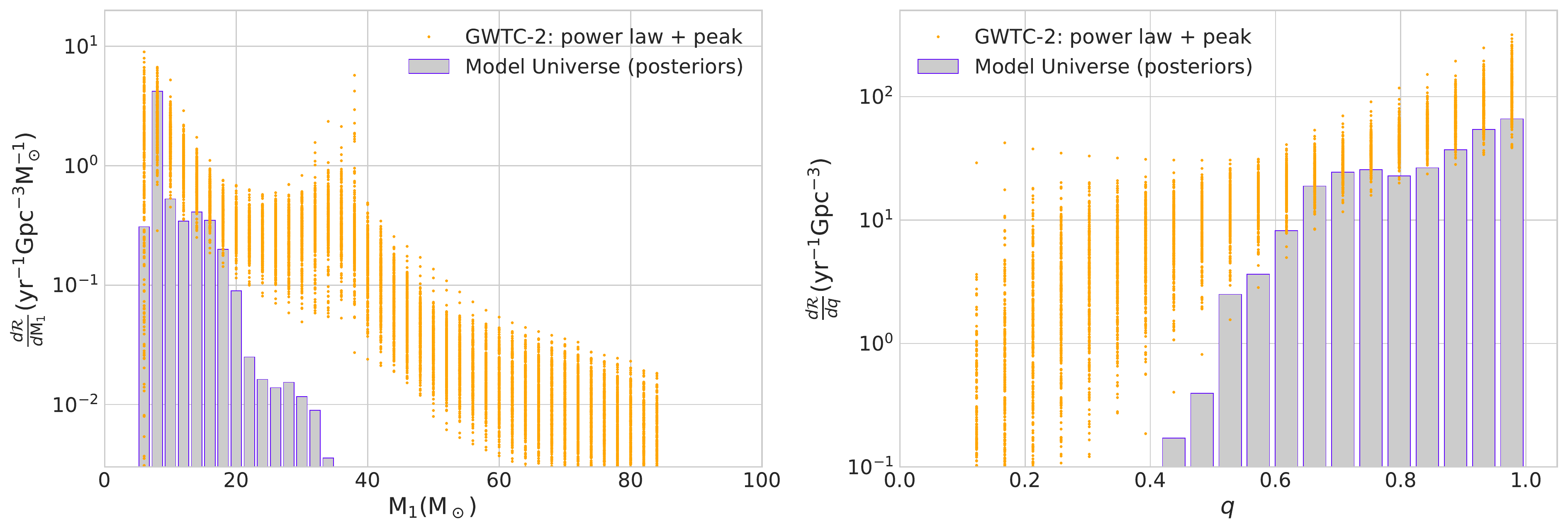}\\
\includegraphics[width=8.0cm,angle=0]{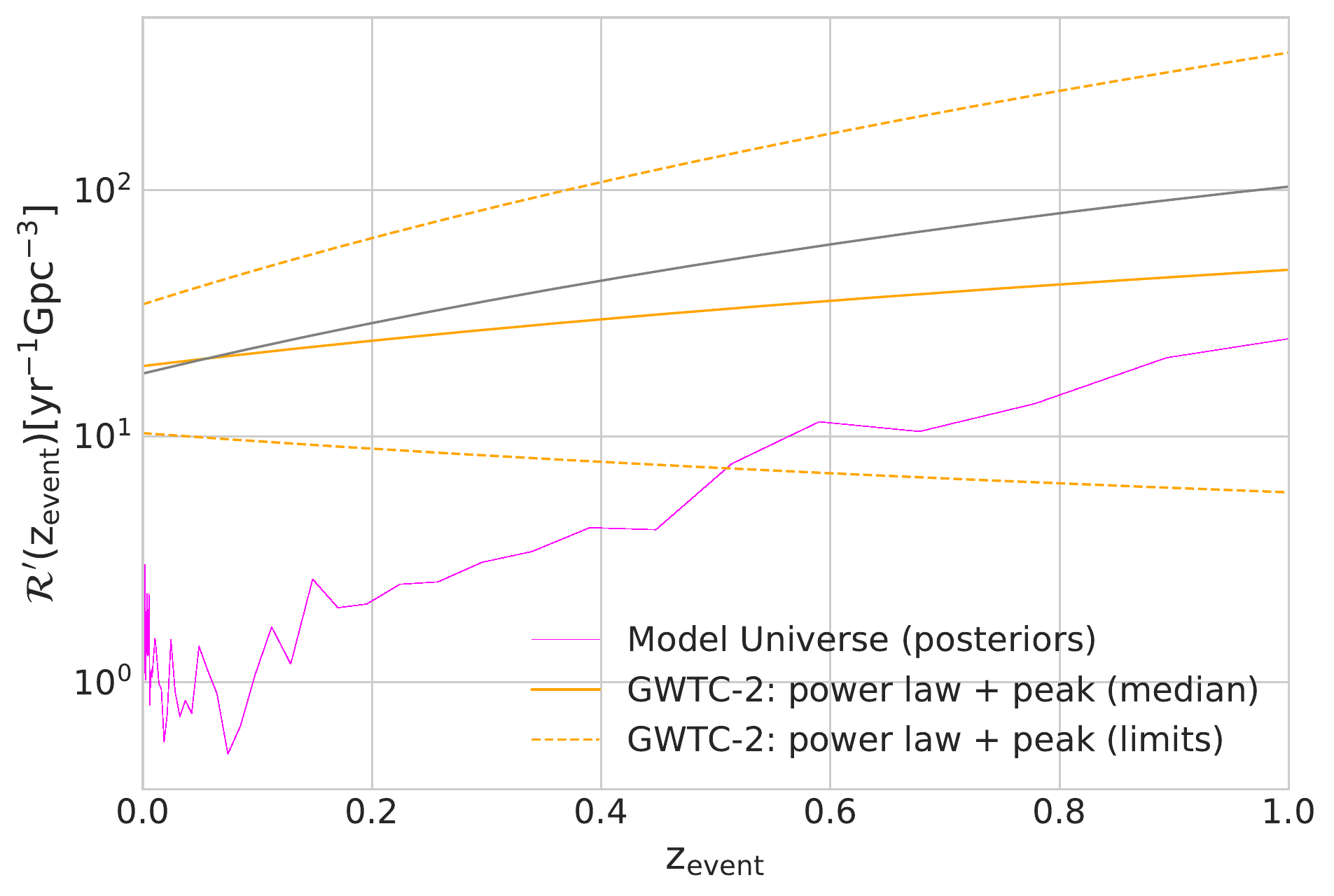}\\
\textbf{1-channel: IB ($\ace=3$)}\par\medskip
\includegraphics[width=15.0cm,angle=0]{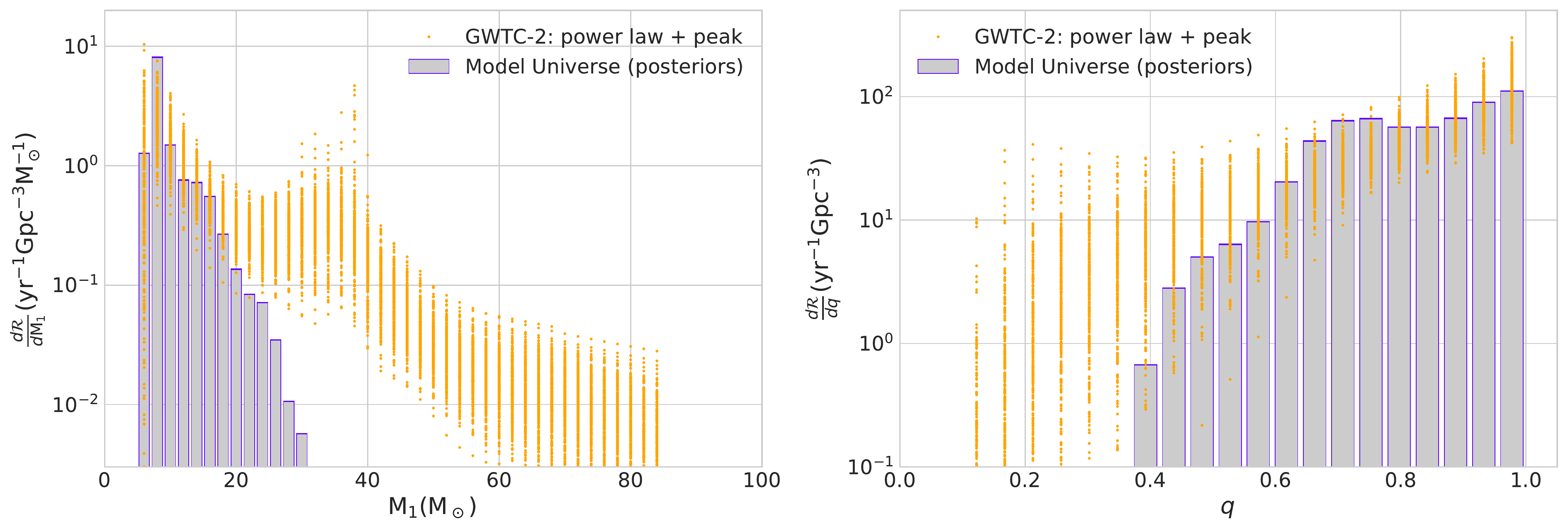}\\
\includegraphics[width=8.0cm,angle=0]{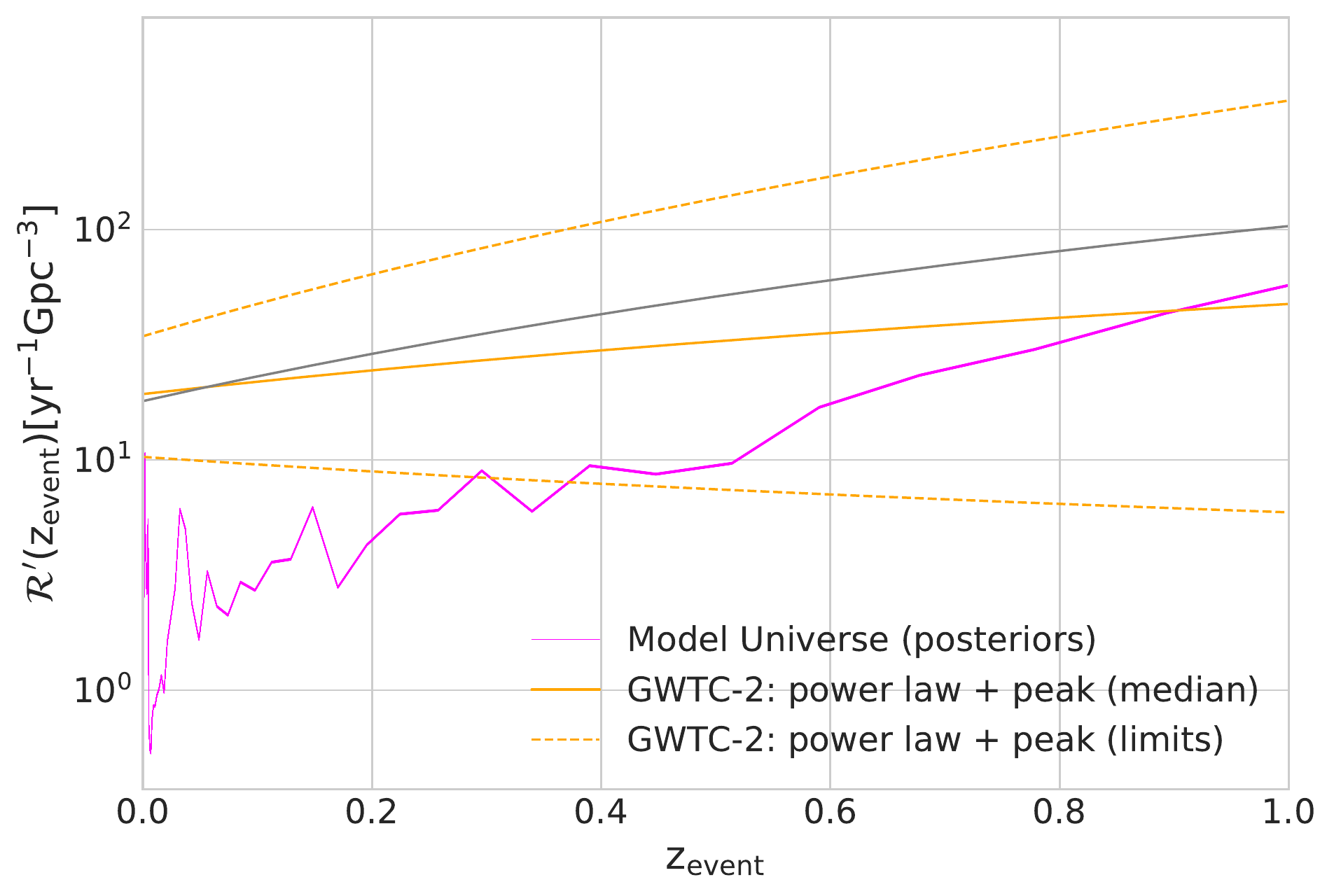}
	\caption{The same description as in Fig.~\ref{fig:rates_1ch_ymc} applies to both of
	the 3-panel sets except for the assumption here that the Model Universe produces
	BBH mergers only due to isolated evolution of massive binaries. Accordingly,
	the posteriors of $\fobin$ in the IB-only universe (Fig.~\ref{fig:post_1ch})
	are applied.
	The cases for $\ace=1$ (top set) and $\ace=3$ (bottom set) are shown.}
\label{fig:rates_1ch_iso}
\end{figure*}

The two sets of panels in Fig.~\ref{fig:rates_1ch_iso} analogously show
the outcomes of the Model Universe with $\fymc=0$, \ie, when only IBs of the universe
produce BBH mergers. The upper (lower) set is for IBs with $\ace=1$ ($\ace=3$). 
Fig.~\ref{fig:rates_1ch_iso} suggests that with mean $\langle\fobin\rangle\approx1.0$
(see Table~\ref{tab_vals})
for $\ace=1$ and $\ace=3$, the Model Universe
$d\rate/d\mone$ falls short of the GWTC-2 rates for $\mone\gtrsim20\Ms$.
The Model Universe $d\rate/dq$, however, well reproduces the corresponding
GWTC-2 differential rates down to $q\approx0.4$, especially with $\ace=3$.
The corresponding $\rz(\zevnt)$ falls below the GWTC-2 lower limit for
$\zevnt\lesssim0.4$, for both $\ace$.

A binary fraction of $\fobin\gtrsim90$\% is consistent with the observed high binary fraction
among OB-stars in clusters and in the Galactic field \citep{Sana_2011,Sana_2013,Moe_2017}.
YMC formation efficiency is a much more ambiguous and poorly determined
quantity \citep{Lada_2003,Longmore_2014,Krumholz_2019}. The inferred
$\fymc\sim10^{-2}$ is consistent with the results from recent
cosmological-hydrodynamical simulations of galaxy and cluster formation  
\citep{Pfeffer_2019}, for upper cutoff of $>10^5\Ms$ of the young cluster mass distribution as
applicable for the present YMC models.

\subsection{Two-channel universe}\label{univ_2ch}

\begin{figure*}
\centering
\textbf{2-channel: YMC/OC + IB ($\ace=1$)}\par\medskip
\includegraphics[width=15.0cm,angle=0]{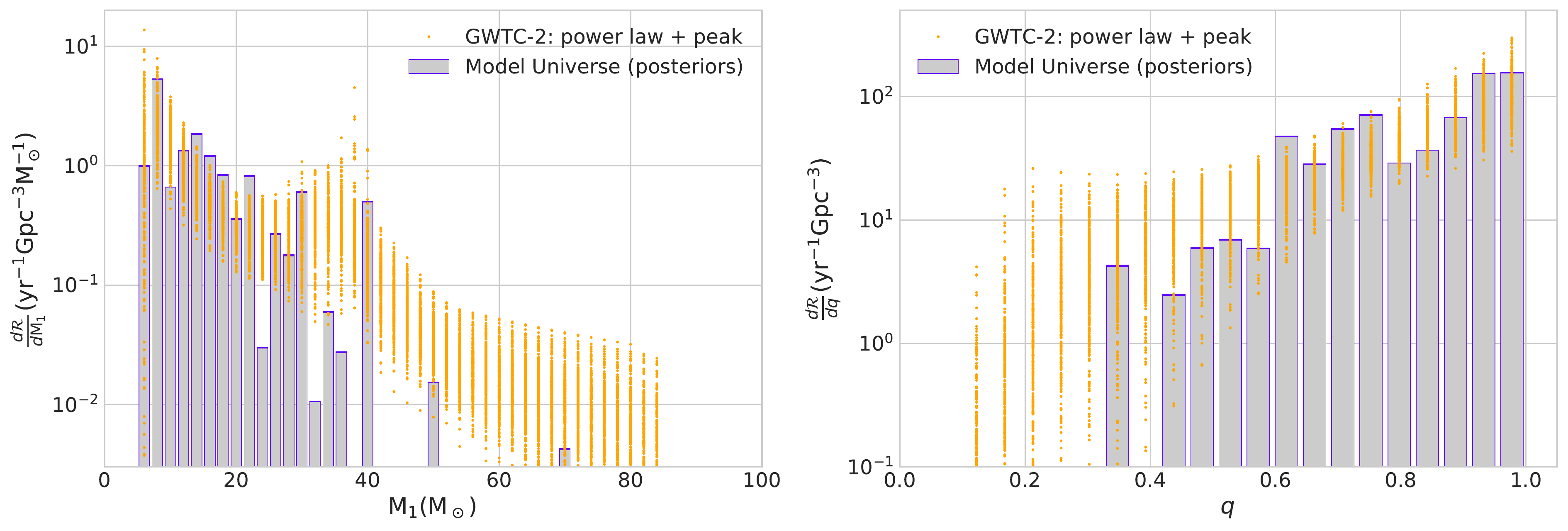}\\
\includegraphics[width=15.0cm,angle=0]{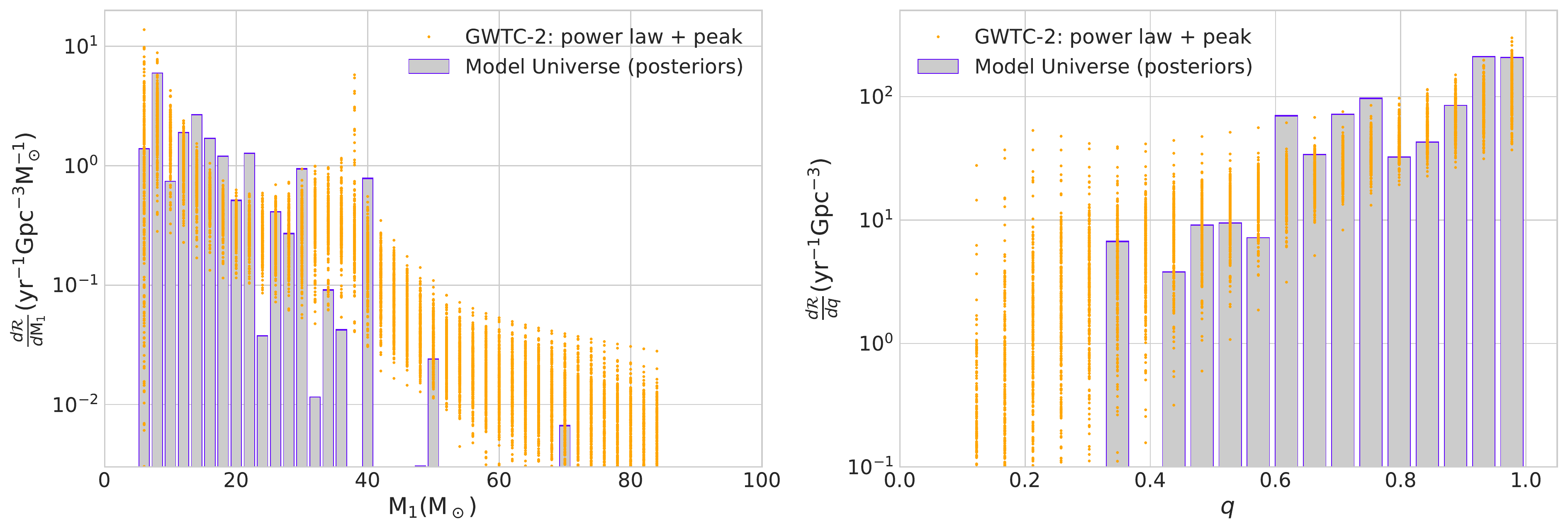}\\
\includegraphics[width=15.0cm,angle=0]{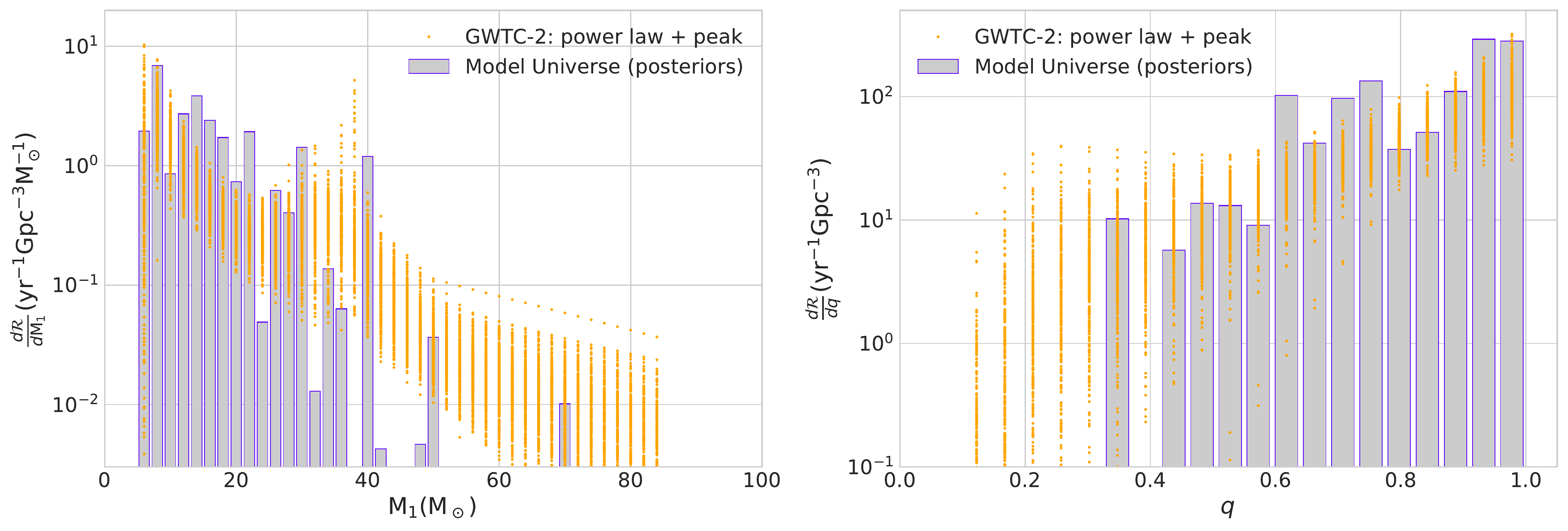}\\
\includegraphics[width=15.0cm,angle=0]{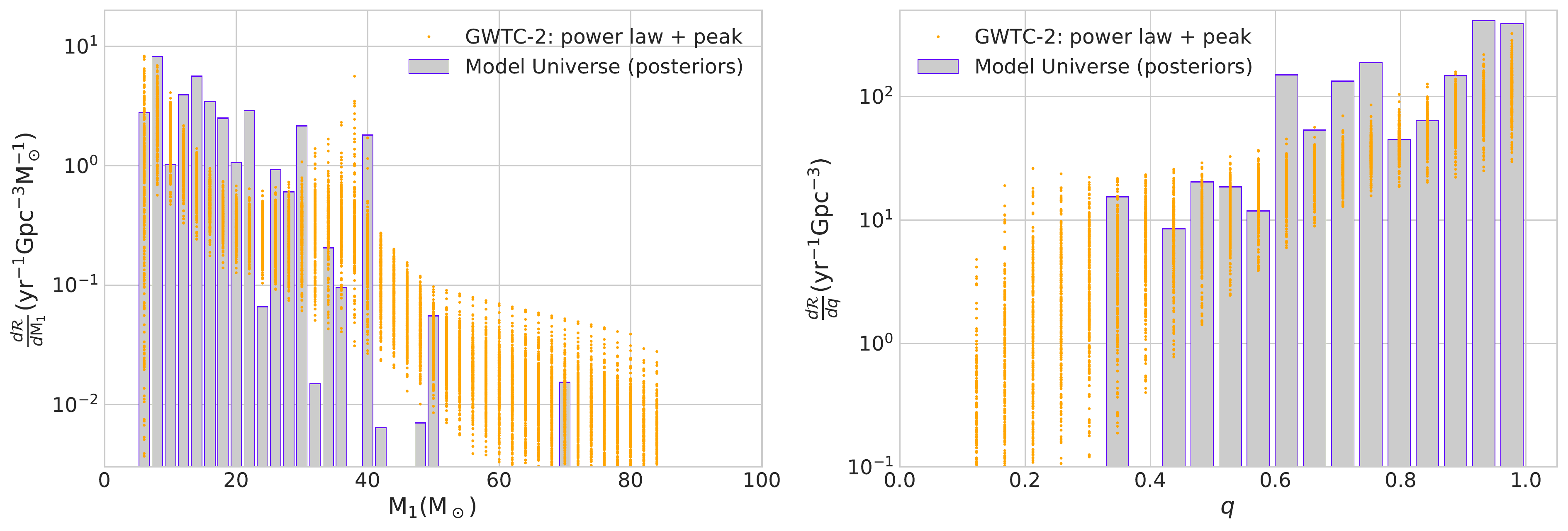}
\caption{The filled histogram is the present-day, differential intrinsic merger rate density of BBHs
	of the Model Universe as obtained by combining those due to the YMC/OC and IB channels
	using the posteriors of $\fymc$ and $\fobin$. The Model Universe posteriors
	are obtained by the two-step Bayesian regression approach described in Sec.~\ref{res}
	(see Fig.~\ref{fig:post_2ch} for an example). Row 1, 2, 3, and 4 corresponds
	to the use of the first, second, third, and fourth moments of $d\rate/d\mone$
	and $d\rate/dq$ (IB evolution with $\ace=1$), respectively. On each panel,
	combined differential merger rate densities for 200 pairs of random and independent
	draws from the posteriors of $\fymc$ and $\fobin$ are superposed.
        As before, the orange dots are random draws (300 per
	bin) of the posteriors of BBH differential intrinsic merger rate densities obtained by
	the LVK GWTC-2 (their power law + peak model).
	}
\label{fig:diffrate_2ch_ace1}
\end{figure*}

\begin{figure*}
\centering
\textbf{2-channel: YMC/OC + IB ($\ace=1$)}\par\medskip
\includegraphics[width=7.5cm,angle=0]{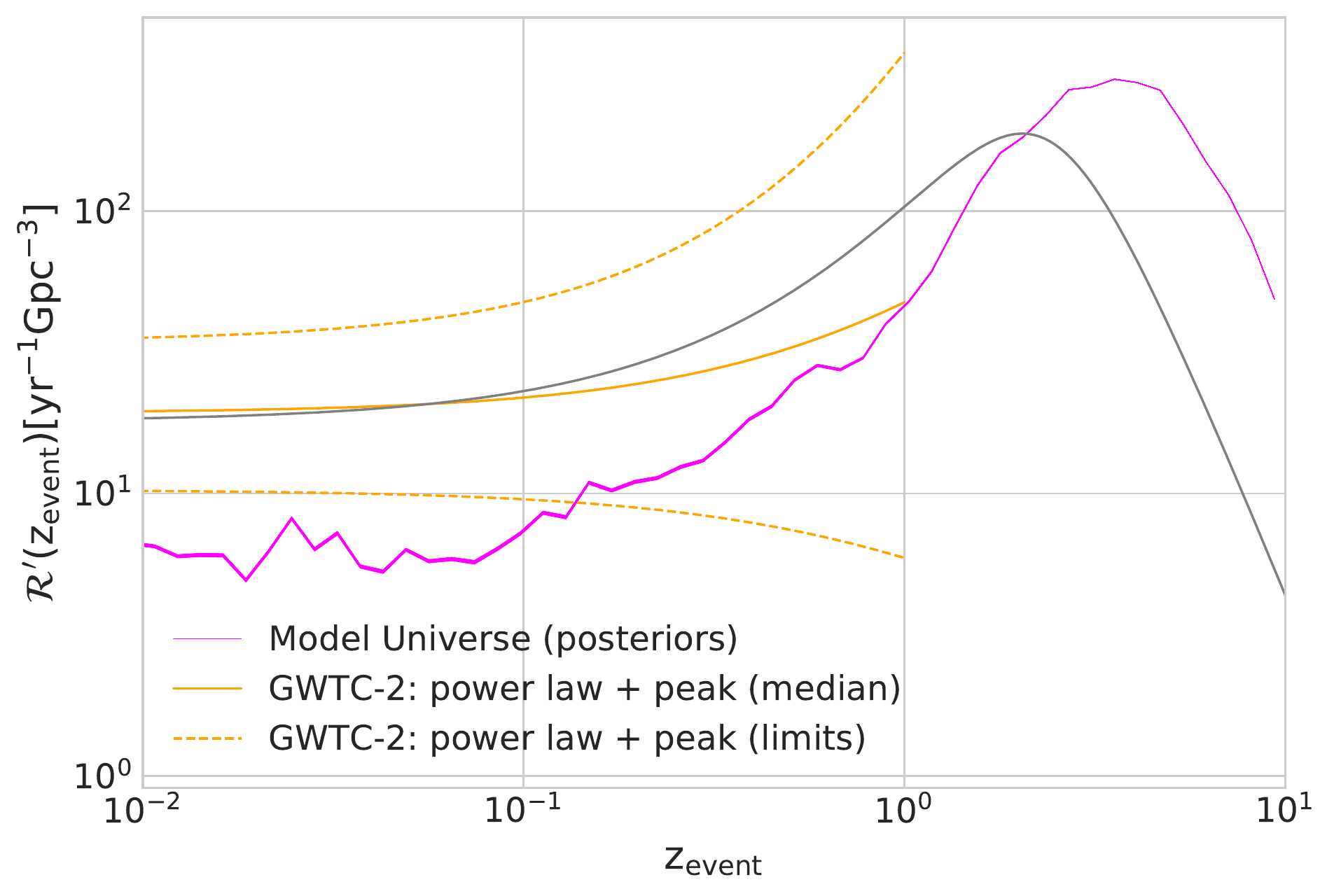}
\includegraphics[width=7.5cm,angle=0]{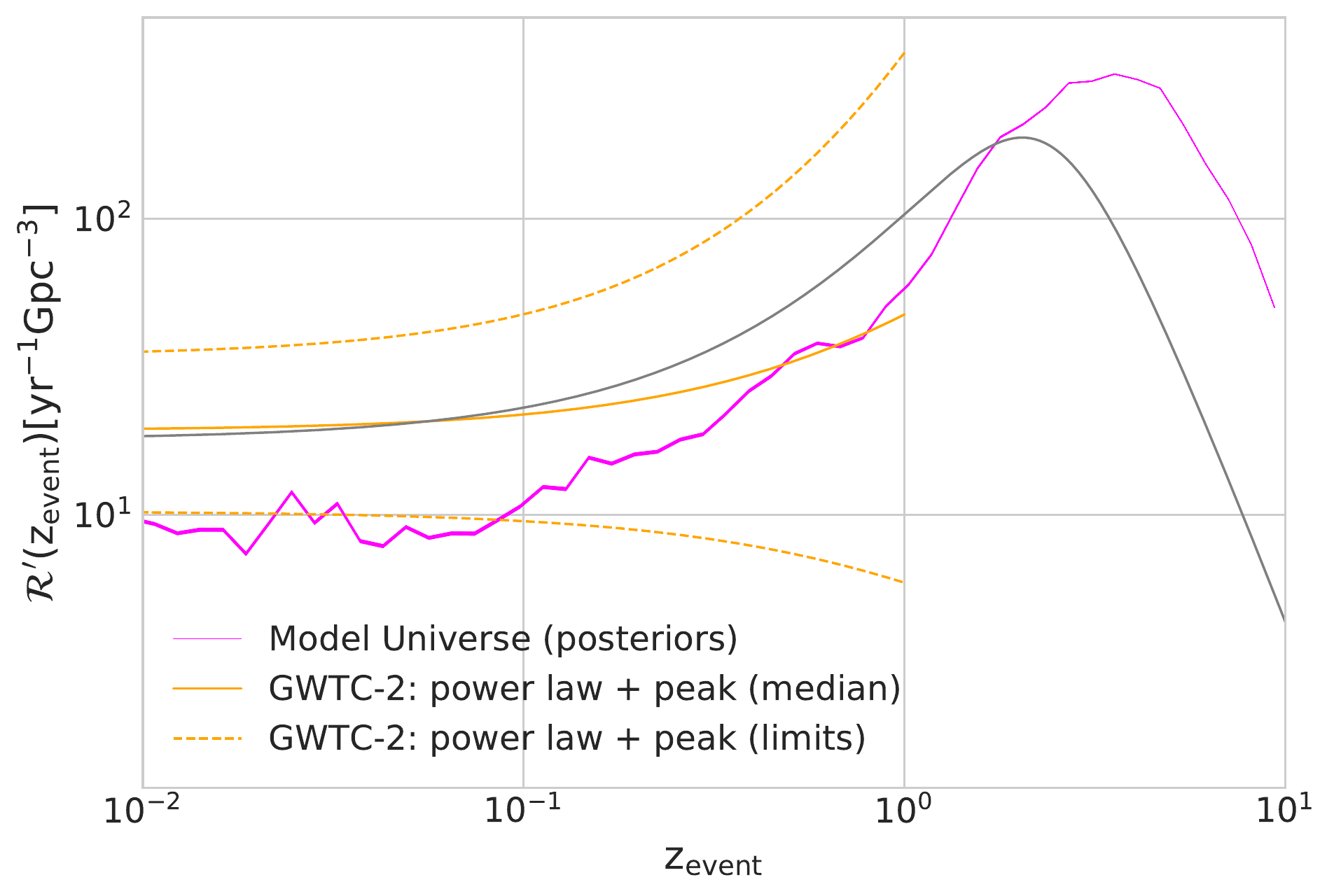}\\
\includegraphics[width=7.5cm,angle=0]{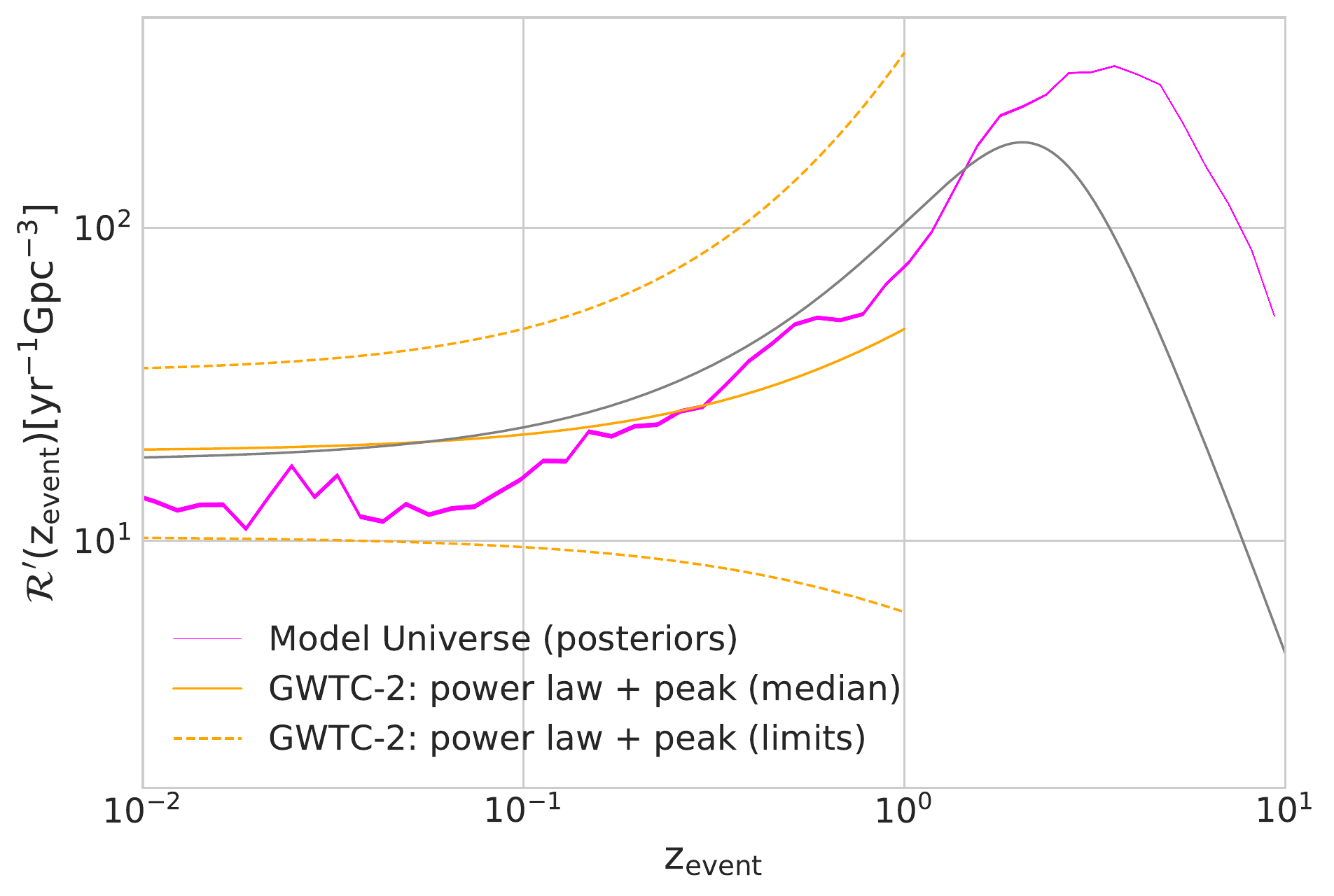}
\includegraphics[width=7.5cm,angle=0]{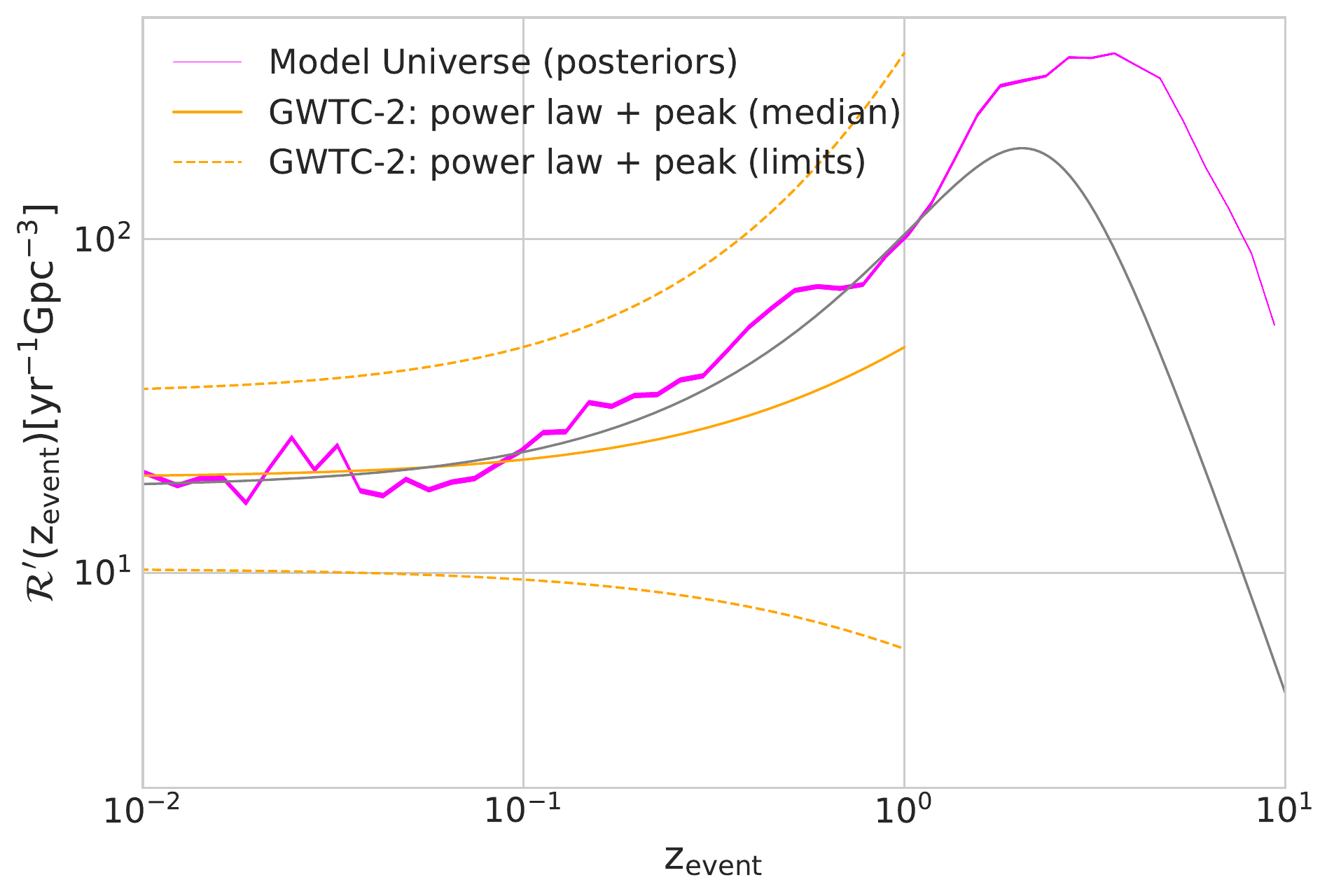}\\
\includegraphics[width=7.5cm,angle=0]{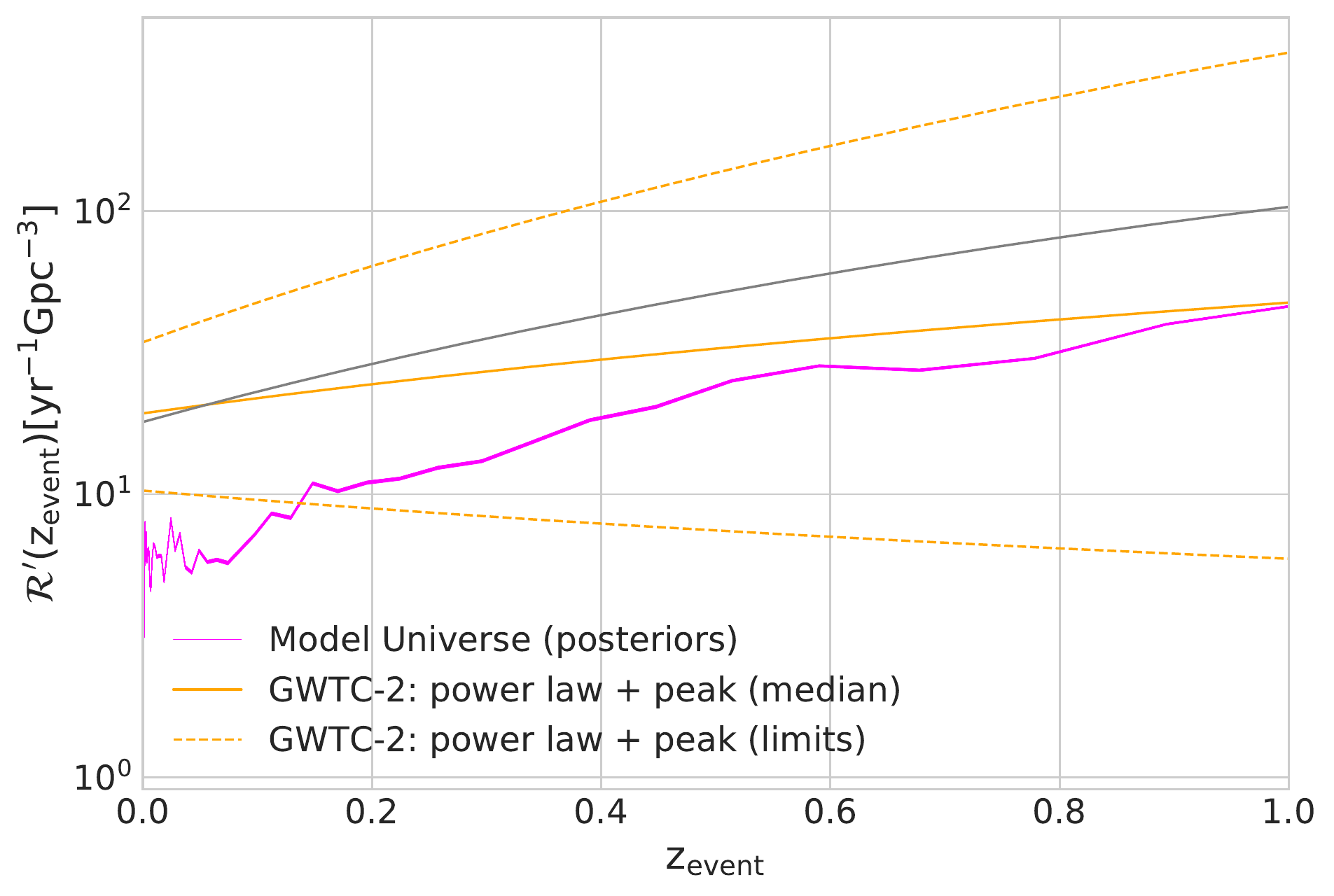}
\includegraphics[width=7.5cm,angle=0]{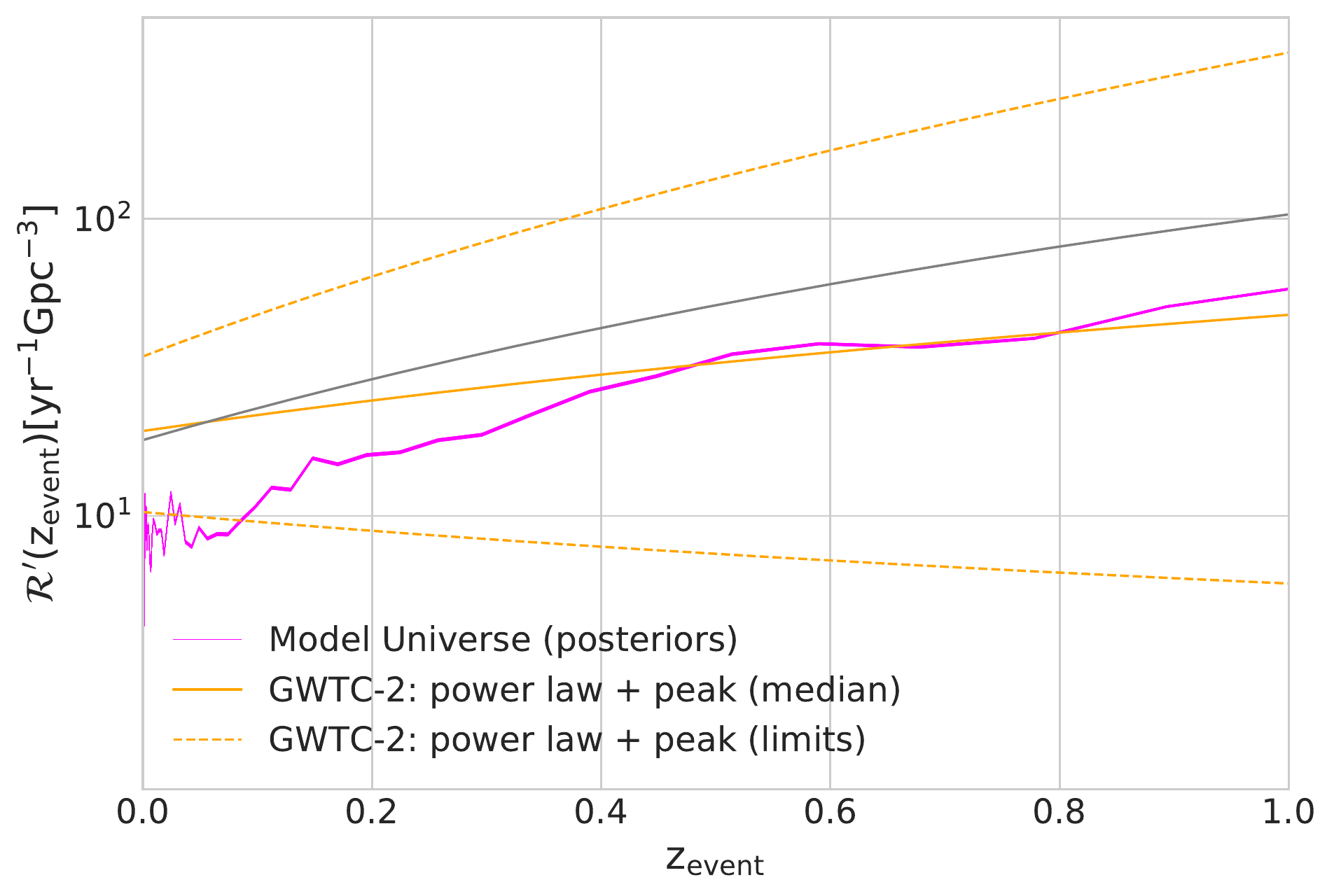}\\
\includegraphics[width=7.5cm,angle=0]{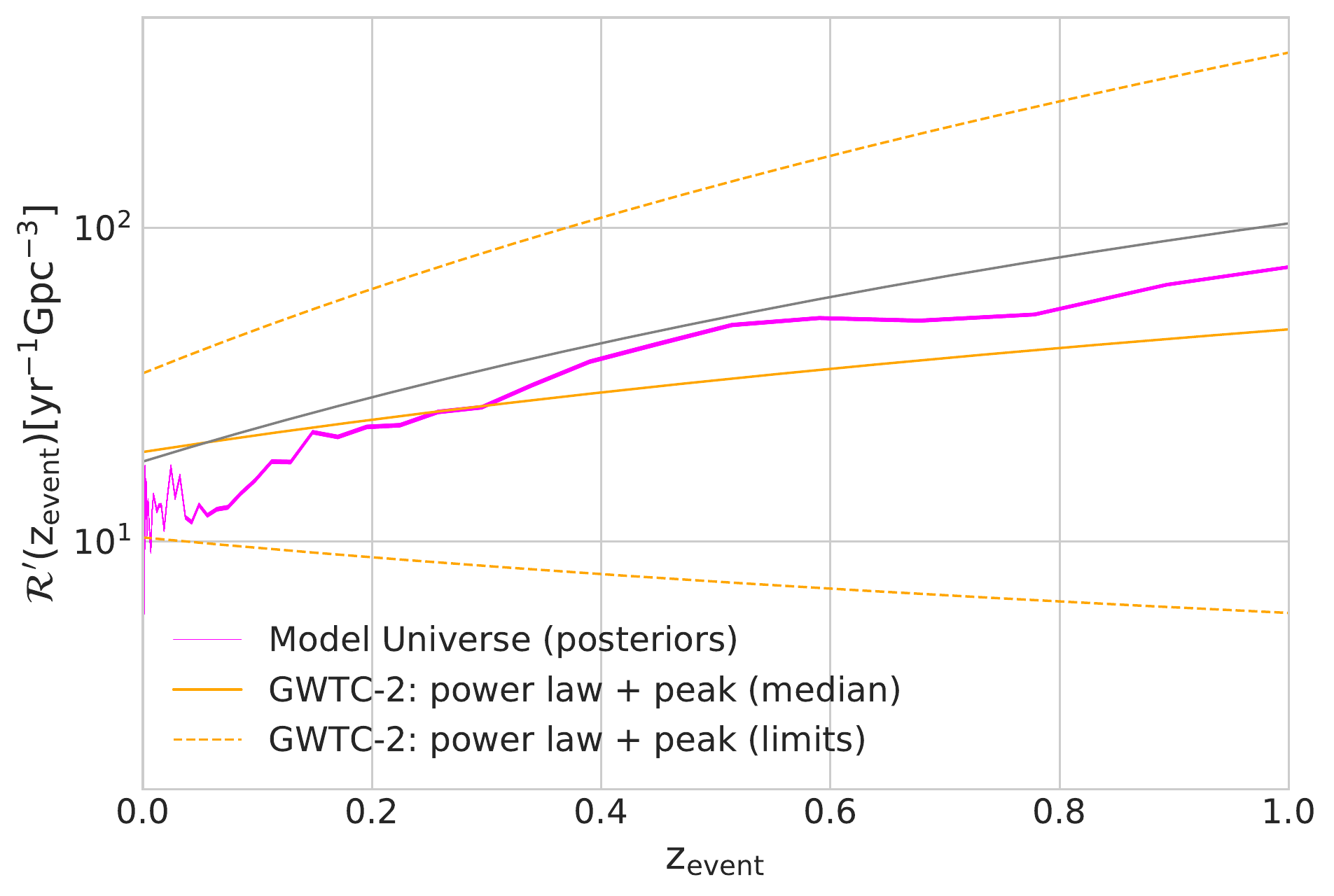}
\includegraphics[width=7.5cm,angle=0]{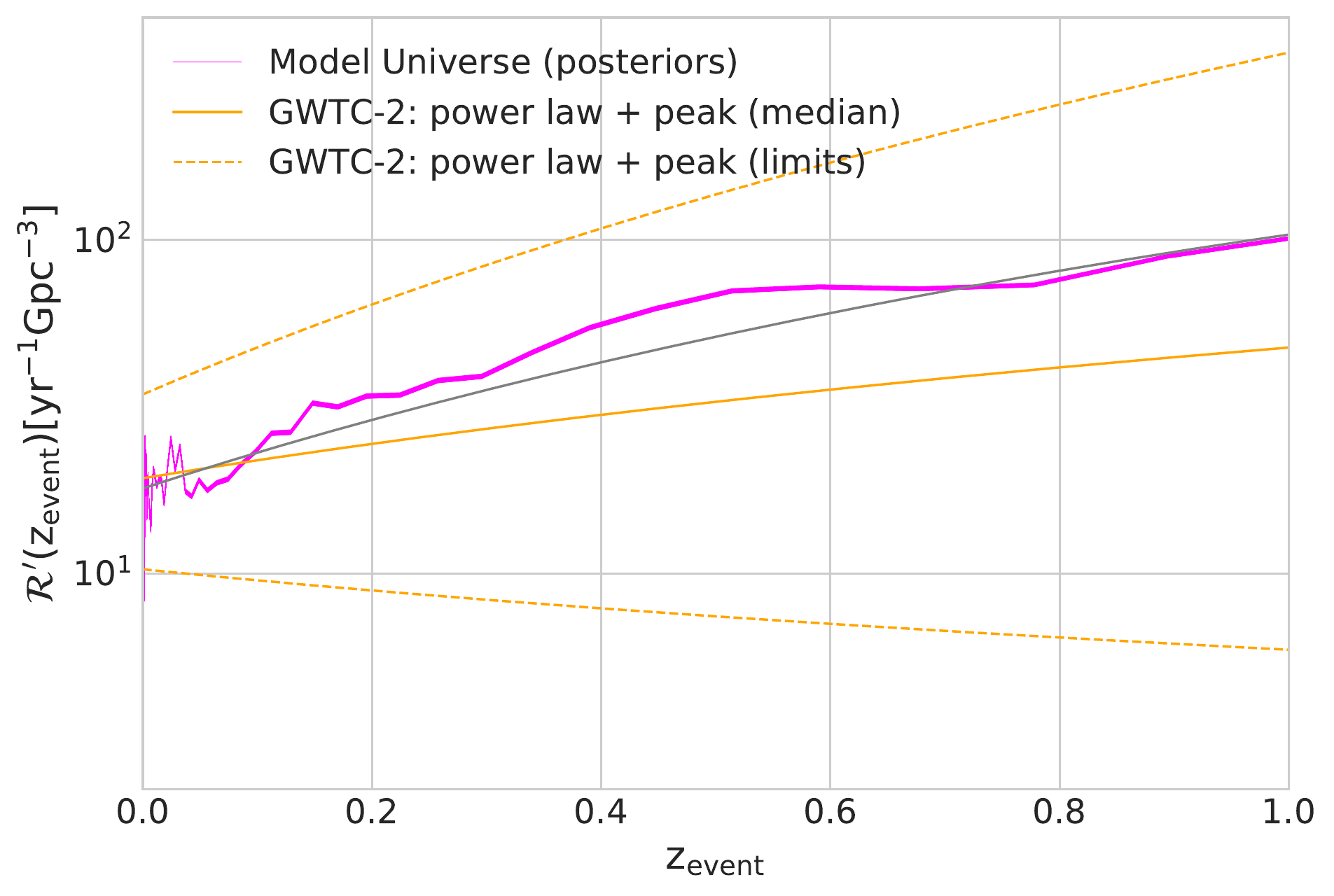}
	\caption{
	The magenta line (all panels) represents the redshift-evolution of
	Model Universe BBH merger rate density from the YMC/OC and IB channels combined.
	As in Fig.~\ref{fig:diffrate_2ch_ace1}, the Model Universe posteriors
	of $\fymc$ and $\fobin$ (200 random and independent pairs of them)
	are applied to combine the rates from the two channels. The first, second, third,
	and fourth panel corresponds
	to the use of the first, second, third, and fourth moments of $d\rate/d\mone$
	and $d\rate/dq$ (IB evolution with $\ace=1$), respectively.
	As in Fig.~\ref{fig:rates_1ch_ymc}, the orange lines depict the corresponding
	GTWC-2 median and 90\% confidence limits (power law + peak model).
	As before, the grey line depicts the
	variation of cosmic SFR with redshift (arbitrary unit along the Y-axis).
	Panels fifth to eighth re-plot these in the same order with the X-axis in linear scale and
	truncated at redshift $1.0$.
	}
\label{fig:rz_2ch_ace1}
\end{figure*}

\begin{figure*}
\centering
\textbf{2-channel: YMC/OC + IB ($\ace=3$)}\par\medskip
\includegraphics[width=13.0cm,angle=0]{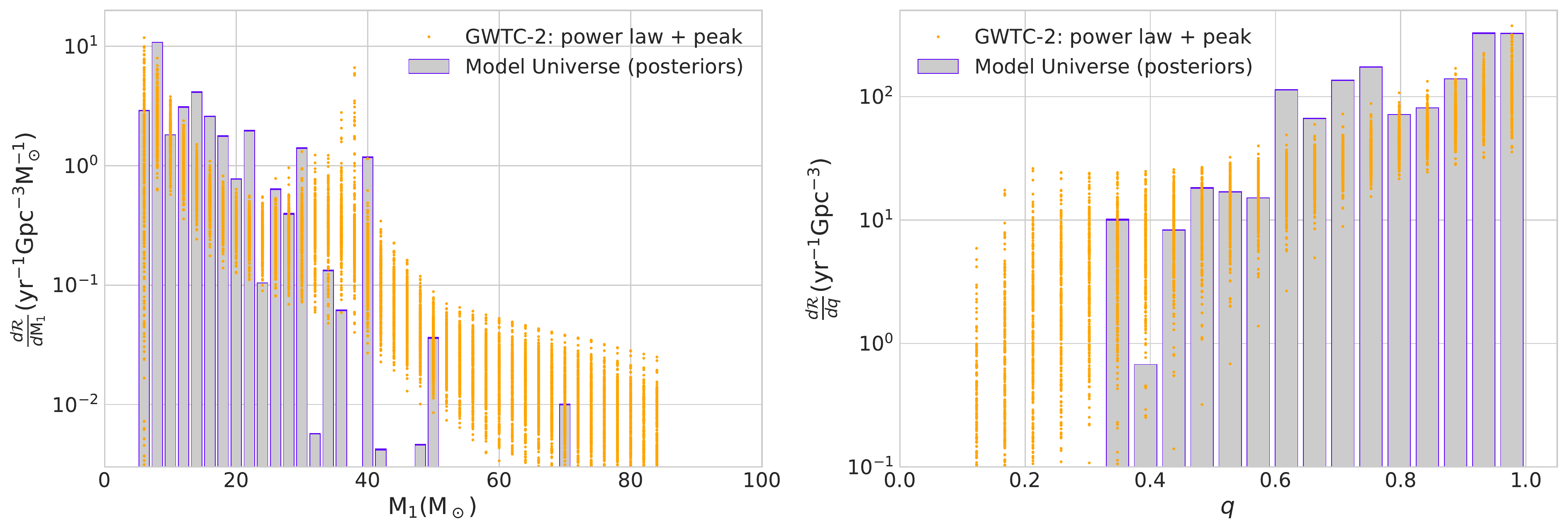}\\
\includegraphics[width=13.0cm,angle=0]{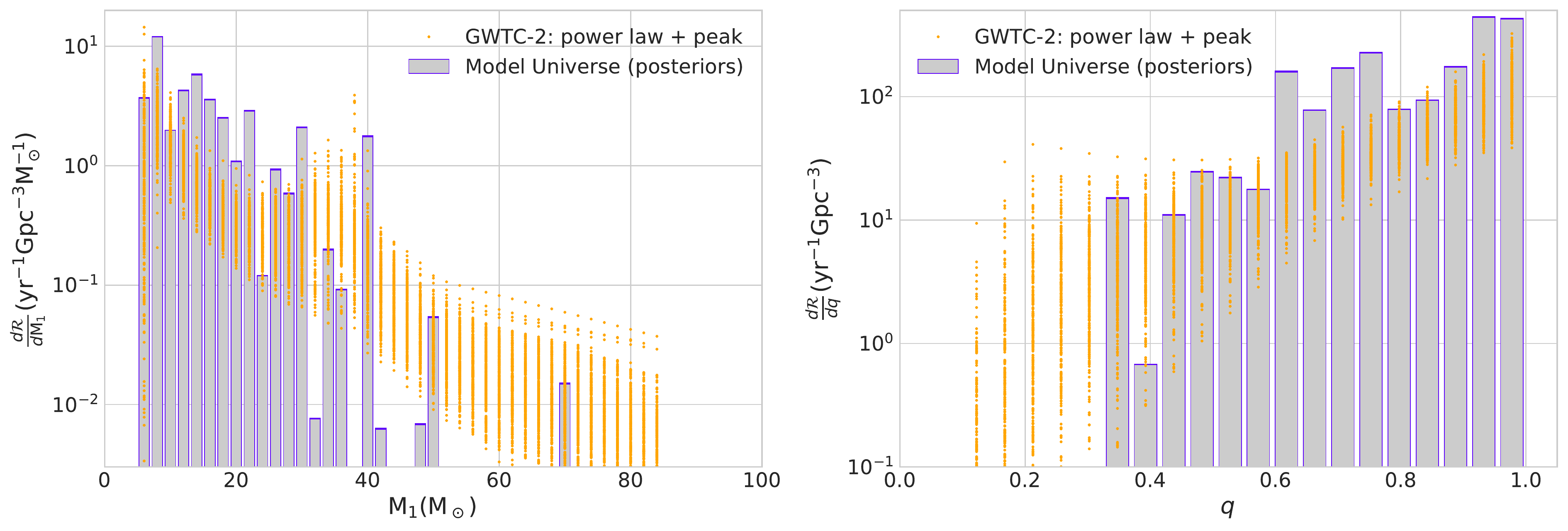}
	\caption{The same description of Fig.~\ref{fig:diffrate_2ch_ace1} applies here
	except that the IB-evolution is with $\ace=3$. The top (bottom) panel corresponds to
	utilizing the third (fourth) moment of the differential rates.}
\label{fig:diffrate_2ch_ace3}
\end{figure*}

\begin{figure*}
\centering
\textbf{2-channel: YMC/OC + IB ($\ace=3$)}\par\medskip
\includegraphics[width=7.0cm,angle=0]{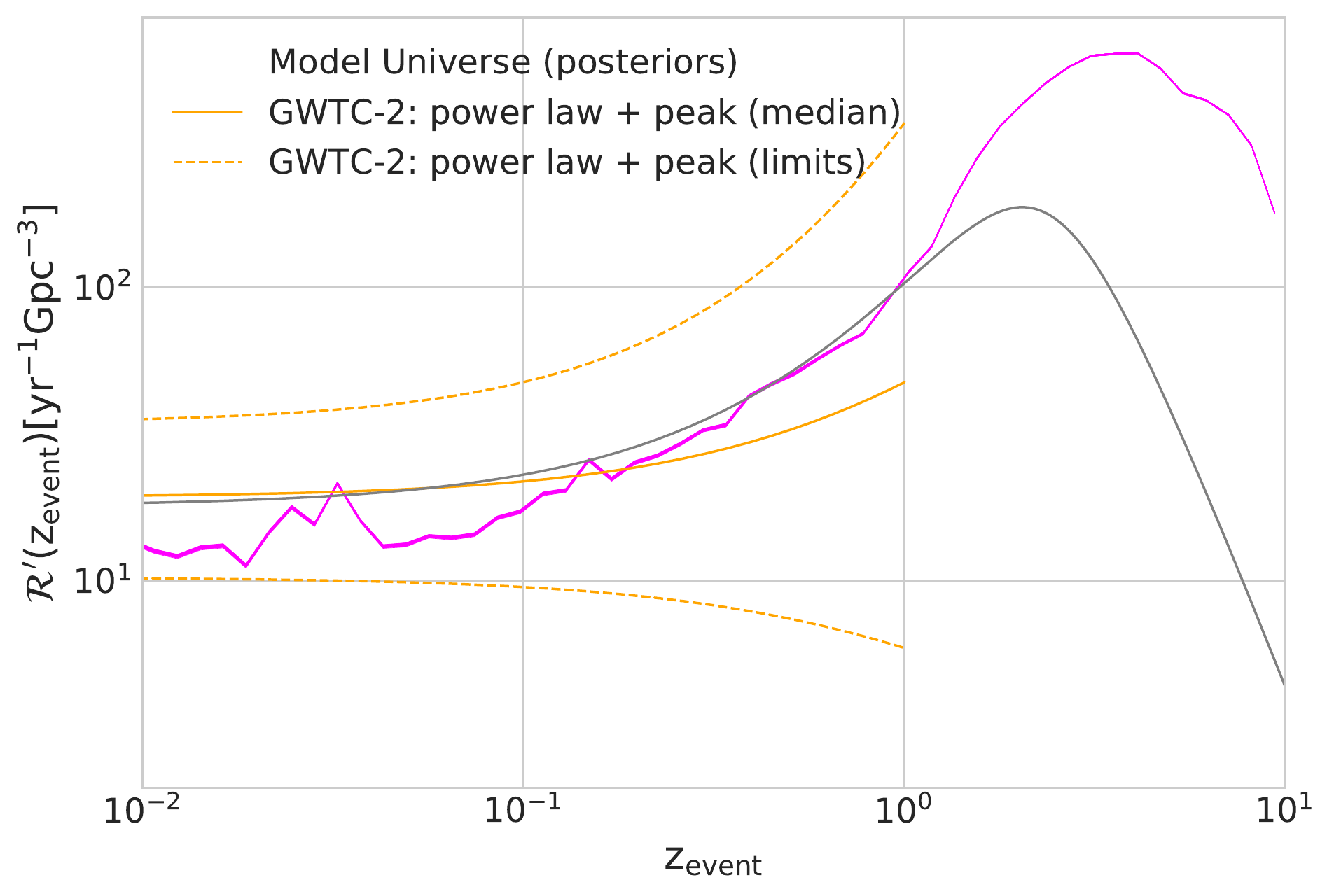}
\includegraphics[width=7.0cm,angle=0]{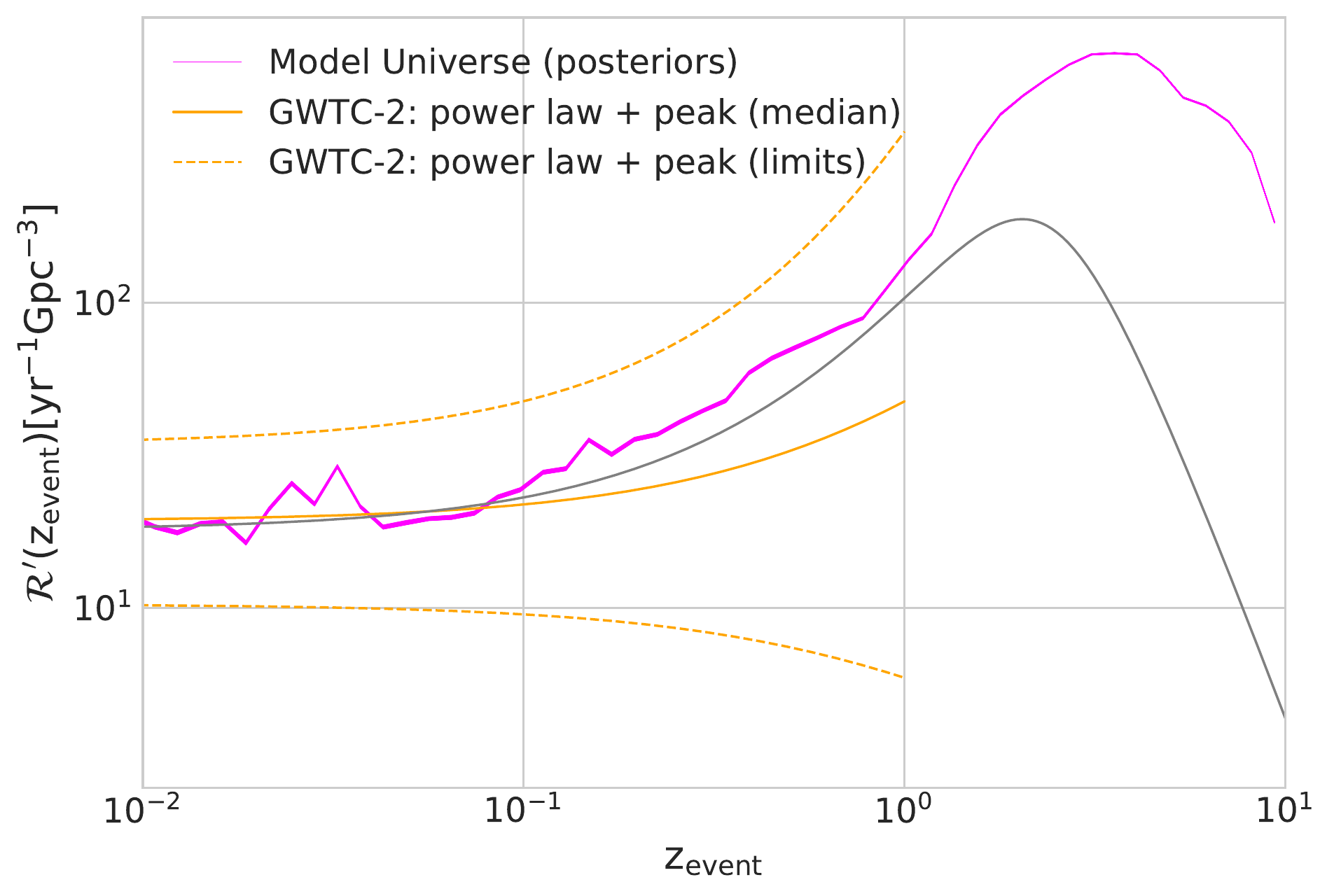}\\
\includegraphics[width=7.0cm,angle=0]{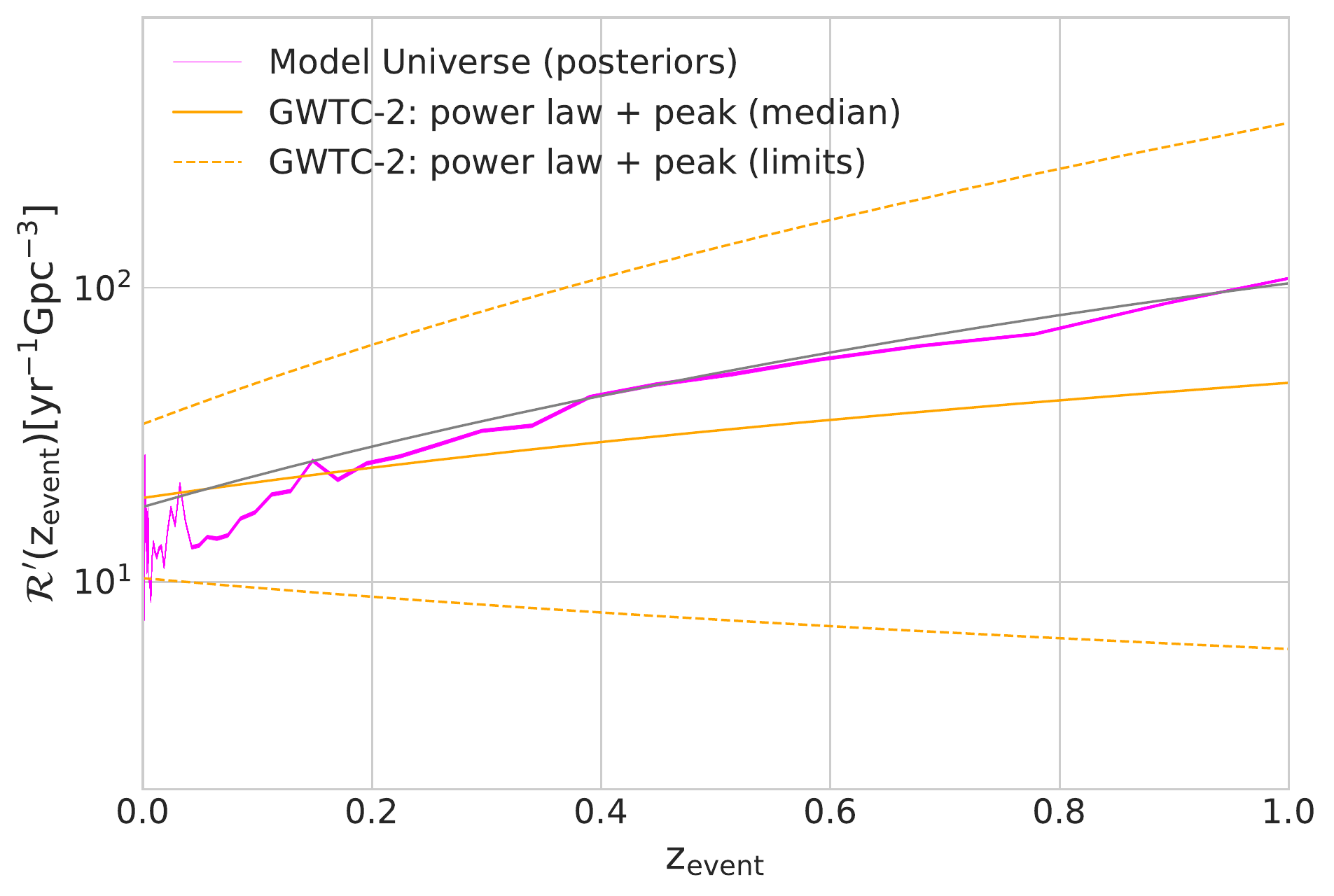}
\includegraphics[width=7.0cm,angle=0]{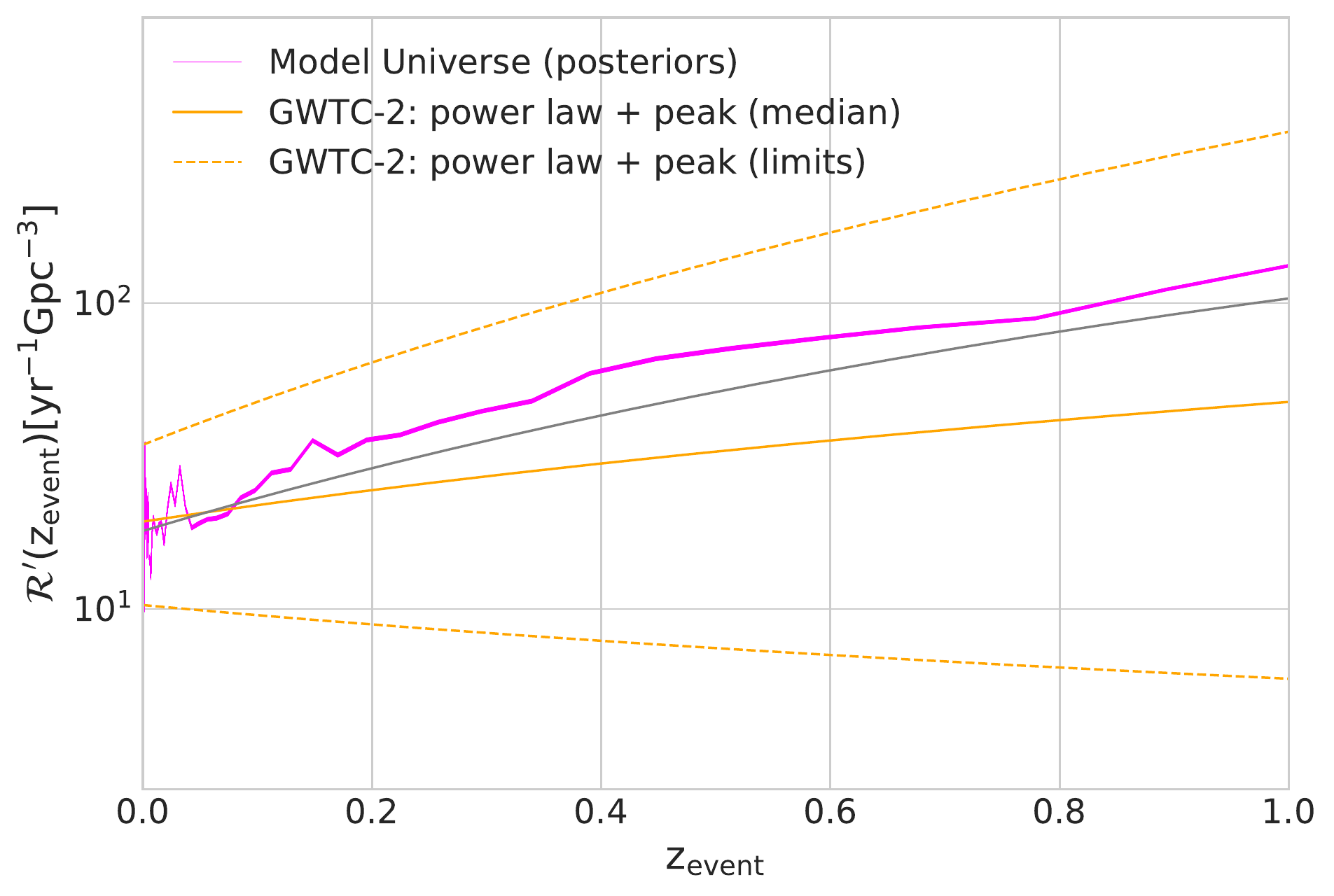}
	\caption{The same description of Fig.~\ref{fig:rz_2ch_ace1} applies here
	except that the IB-evolution is with $\ace=3$. The upper-left (-right) panel
	corresponds to utilizing the third (fourth) moment of the differential rates.
	The lower panels re-plot these in the same order with the X-axis in linear scale and
	truncated at redshift $1.0$.
	}
\label{fig:rz_2ch_ace3}
\end{figure*}

Fig.~\ref{fig:diffrate_2ch_ace1} shows the combined $d\rate/d\mone$
and $d\rate/dq$ for the two-channel Model Universe where BBH mergers are
produced both dynamically in YMC/OCs and through IB evolution (Sec.~\ref{res}).
The Model Universe differential rates are shown for the Bayesian regression
analyses using moments of order $p=1$, 2, 3, and 4 of
the rate distributions (Sec.~\ref{res}) and taking $\ace=1$ for the IB evolution. 
Fig.~\ref{fig:rz_2ch_ace1} shows the corresponding combined $\rz(\zevnt)$ evolutions
(up to both $\zevnt=10$ in logarithmic scale and $\zevnt=1$ in linear scale).
Fig.~\ref{fig:diffrate_2ch_ace3} and \ref{fig:rz_2ch_ace3} show the
Model Universe combined $d\rate/d\mone$, $d\rate/dq$, $\rz(\zevnt)$ 
yields when the IBs in the Model Universe evolve with $\ace=3$;
the results with $p=3$ and 4 are shown. 

With moments of increasing order included in the analysis, $\langle\fymc\rangle$
increases whereas $\langle\fobin\rangle$ stays nearly constant
at $\approx1.0$ ($\approx0.9$) for $\ace=1$ ($\ace=3$); see Table.~\ref{tab_vals}.
This results in overall increase of the present-day, combined differential rates
and of the combined total rate at lower redshifts ($\zevnt\lesssim1$). The high redshift
behaviour of $\rz$ is still dominated by the contribution from IBs (for both $\ace$)
so that the combined $\rz$ is peaked much earlier
(at $\zevnt\approx4$) than the cosmic SFH, similarly as $\rzib$. With  
$\langle\fobin\rangle\gtrsim0.9$ and $\langle\fymc\rangle\lesssim 10^{-2}$,
$\rz$ would be dominated by $\rzib$ at high redshifts as Fig.~\ref{fig:rz_pure} suggests
(see also discussions in Sec.~\ref{popsynth}).

The increase of $\langle\fymc\rangle$ with moment-order, $p$, is due to the
fact that higher order moment would amplify the dependence on larger
$\mone$, where YMC/OCs contribute to the GWTC-2 rates essentially solely.
As seen in Fig.~\ref{fig:diffrate_pure},
for both $\ace$, the profile of $d\riso/d\mone$ declines steeply from
$\mone\gtrsim20\Ms$ and cuts off at $\mone\approx30\Ms$. As opposed to this,
the $d\rymc/d\mone$ profile continues much more smoothly up to $\approx50\Ms$
and also contains discrete events beyond, in the PSN gap (the PSN-gap
BHs being produced via either first-generation BBH mergers or BH-Thorne-Zytkow-Object accretion;
see Ref.~\cite{Banerjee_2020c} for details). On the other hand,  
$d\riso/dq$ already fits well (for both $\ace$; see Fig.~\ref{fig:diffrate_pure}),
without any scaling, the corresponding GWTC-2 rates
over $q\gtrsim0.5$ where most of the total rate is accumulated. Hence,
$\langle\fobin\rangle$ is essentially `settled' from the $q$
dependence of the GWTC-2 and IB rates. As discussed in Sec.~\ref{res},
it is these features of the Model Universe differential rates vis-\'a-vis
those from GWTC-2 that motivates the two-stage Bayesian regression
applied in the 2-channel case.  

With $p=3$ and 4, the two-channel
Model Universe $d\rate/d\mone$, $d\rate/dq$, and $\rz(\zevnt)$ all
agree well with those from GWTC-2 over the relevant ranges of
$\mone$, $q$, and $\zevnt$, as seen in Figs.~\ref{fig:diffrate_2ch_ace1},
\ref{fig:rz_2ch_ace1}, \ref{fig:diffrate_2ch_ace3}, \ref{fig:rz_2ch_ace3}.
This is as opposed to the one-channel universe where the agreements
are partial (see Sec.~\ref{univ_1ch}) despite similar estimated
values of $\fymc$ and $\fobin$ (see Table~\ref{tab_vals}).

All the Model Universe rates, obtained so far, have small variance.
This is due to the small, of the order of Poisson error, variance
that goes into the likelihood functions, Eqn.~\ref{eq:lhood}
(see Sec.~\ref{res}). Introducing astrophysical uncertainties
in the ingredients of the Model Universe would increase the uncertainties
in its rates, as demonstrated here by altering the redshift-metallicity
relation.

The population synthesis exercises of IB-only and YMC-only universes
(Sec.~\ref{popsynth}) are repeated also with the `low-Z' and `high-Z'
cosmic metallicity evolutions \citep{Chruslinska_2019}. The two-channel
universe is then rerun with $\Delta\mupi$s recalculated similarly (Sec.~\ref{res})
but after stacking outcomes, in equal numbers, from the `low-Z', `moderate-Z', and `high-Z'
trials. The $\mupi$s used in the run are also equal-weighted
average of the $d\ratei/dX$s obtained from the three redshift-metallicity dependencies.

\begin{figure*}
\centering
\textbf{Pure channel: YMC/OC and IB ($\ace=3$)}\par\medskip
\includegraphics[width=15.0cm,angle=0]{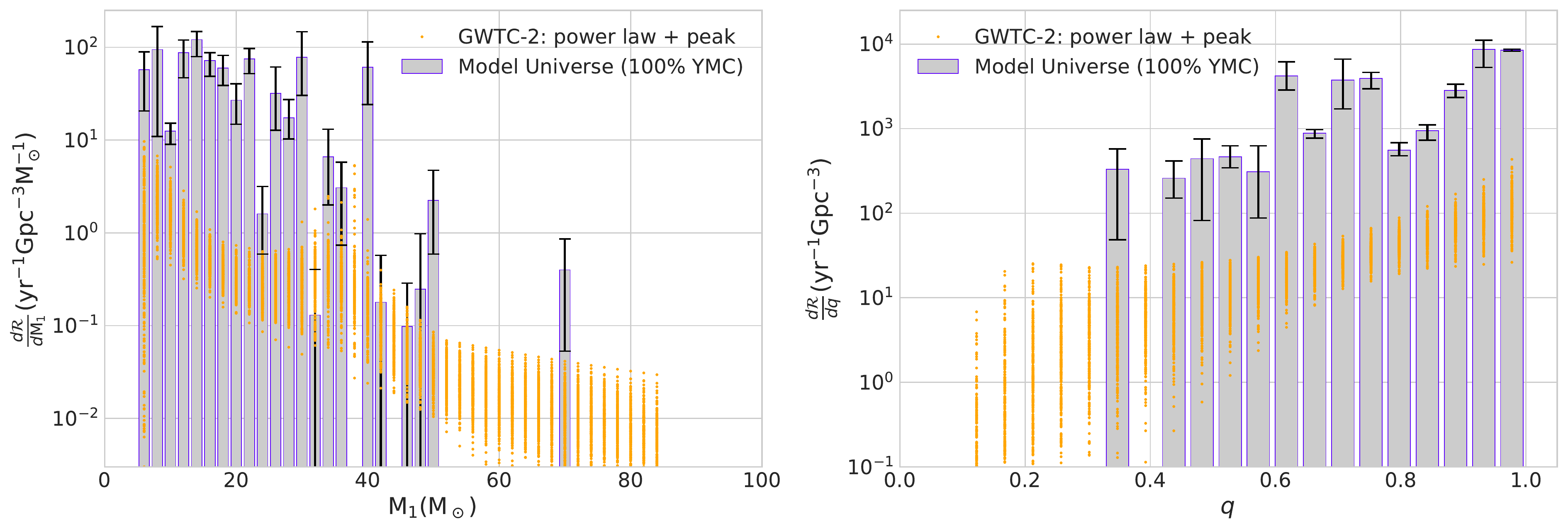}\\
\includegraphics[width=15.0cm,angle=0]{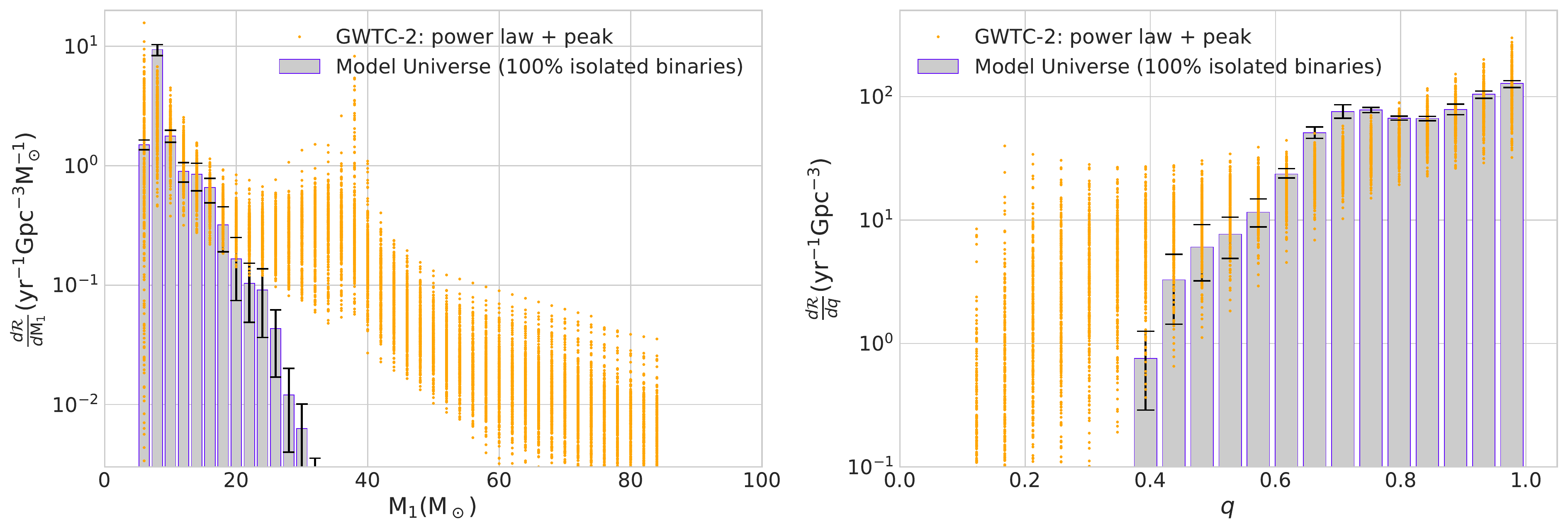}
	\caption{The same description as for Fig.~\ref{fig:diffrate_pure} applies. For
	the IB channel (lower panels), only the $\ace=3$ case is shown. The error bars
	are due to the three different variants of cosmic metallicity evolution history in
	Ref.~\cite{Chruslinska_2019}, namely, their `low-Z', `moderate-Z', and `high-Z'
	evolutions (equal weights applied).}
\label{fig:diffrate_pure2}
\end{figure*}

\begin{figure*}
\centering
\textbf{2-channel: YMC/OC + IB ($\ace=3$)}\par\medskip
\includegraphics[width=15.0cm,angle=0]{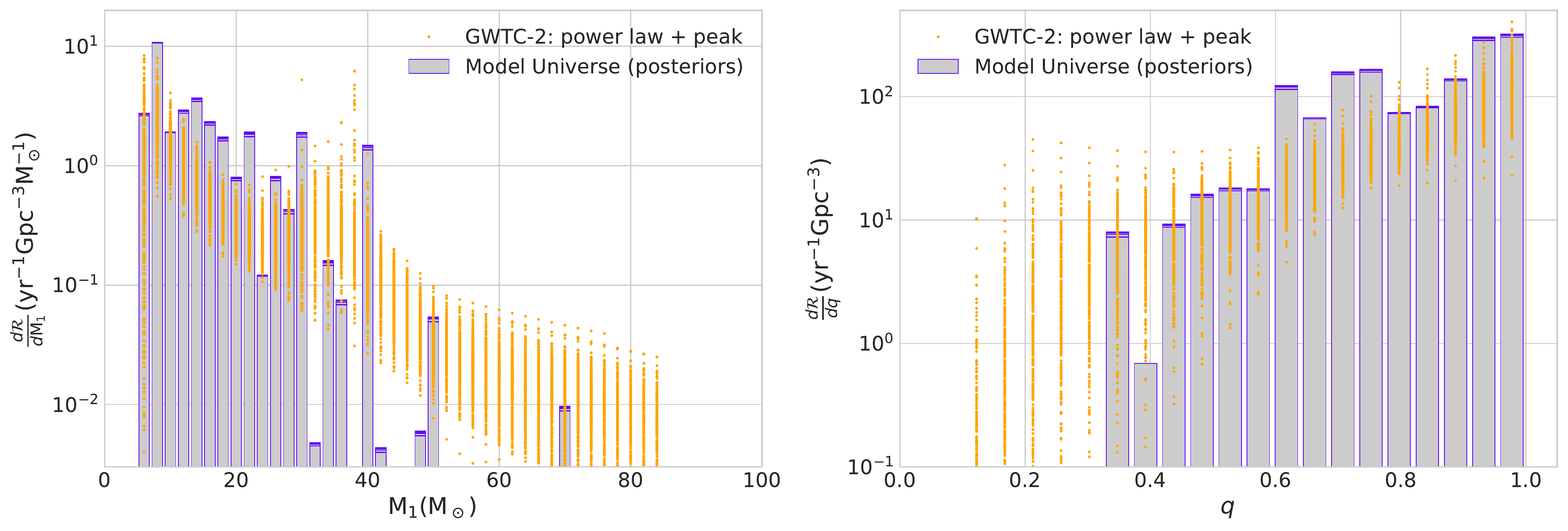}\\
\includegraphics[width=7.5cm,angle=0]{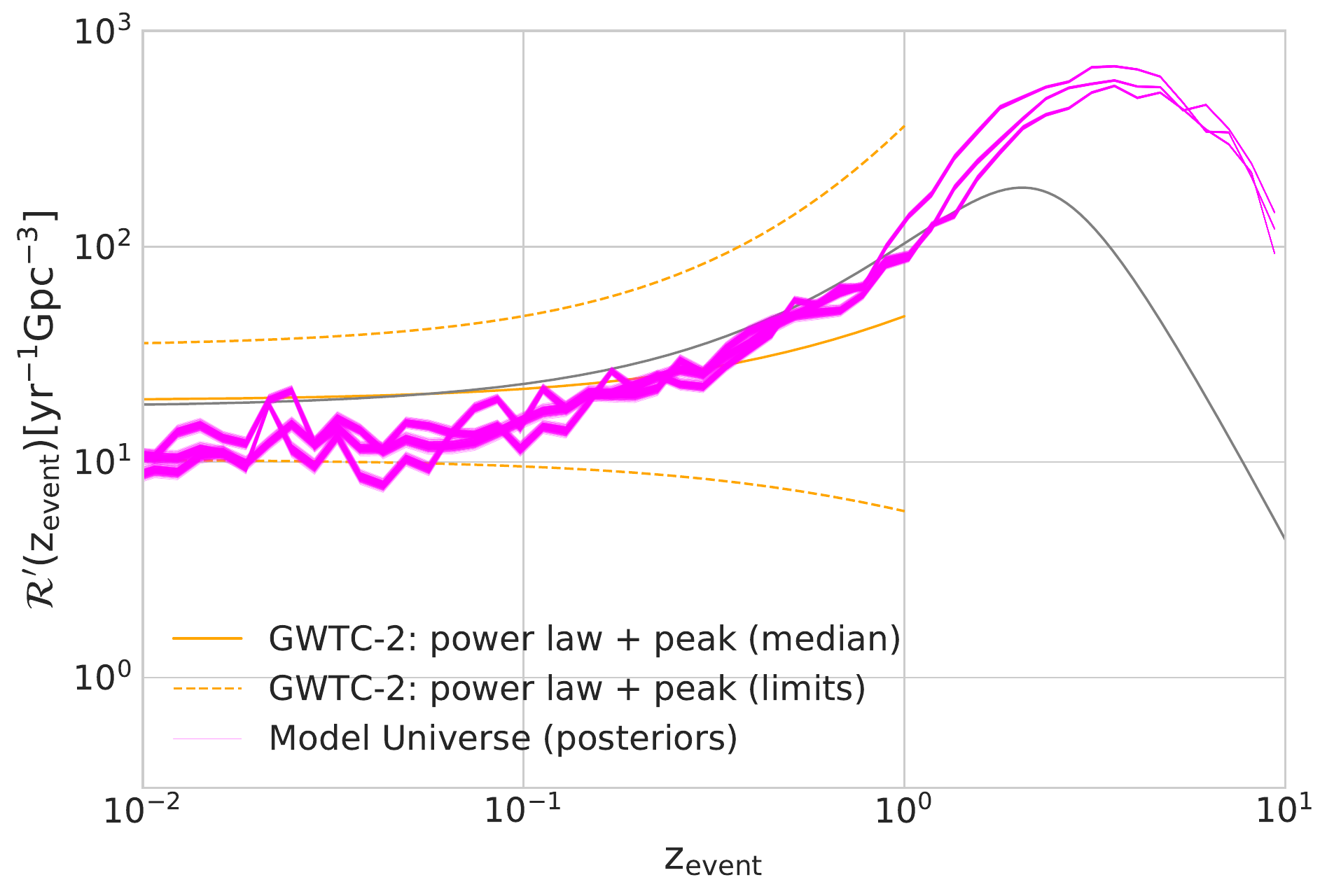}
\includegraphics[width=7.5cm,angle=0]{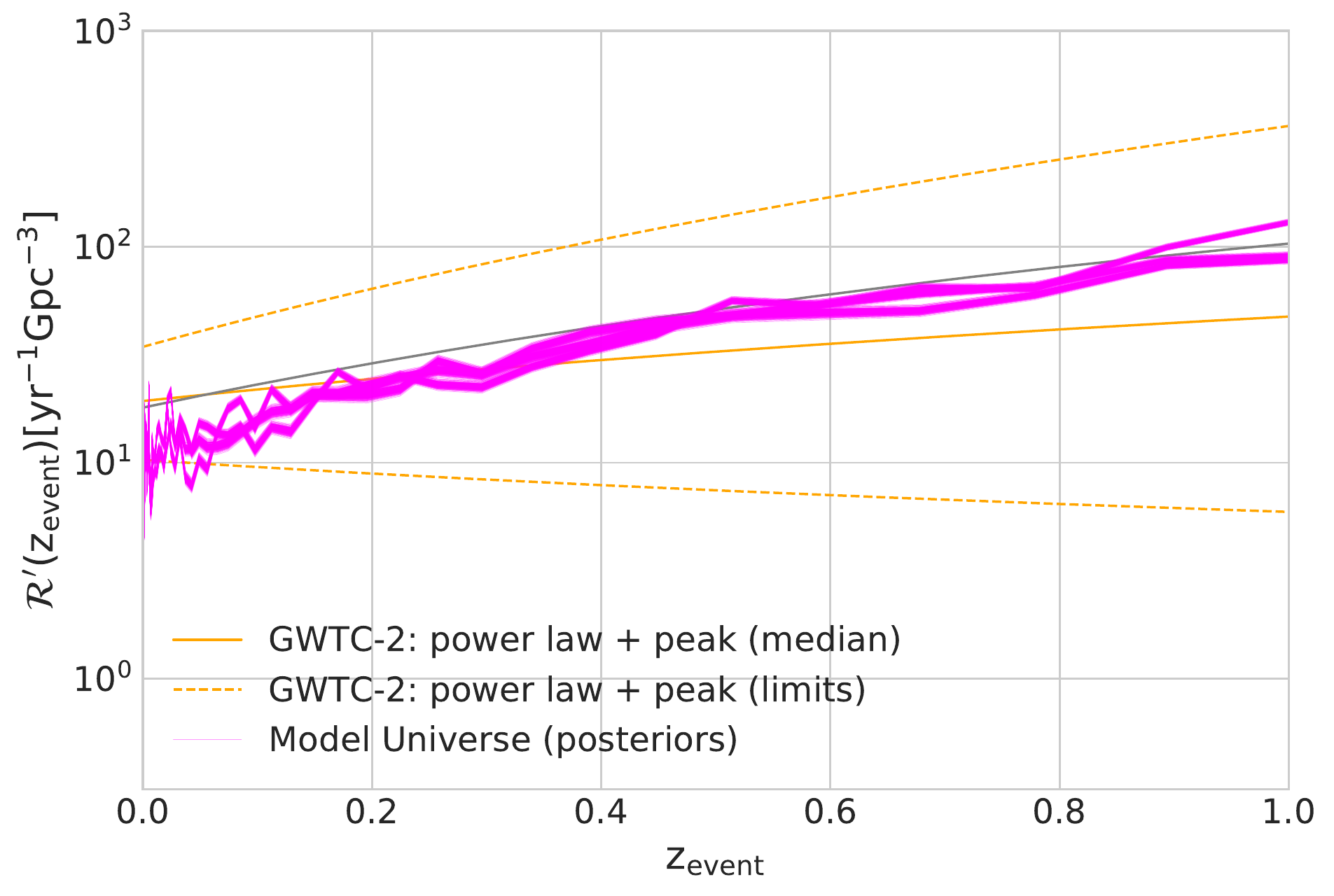}
	\caption{The combined, YMC/OC + IB ($\ace=3$) present-day differential merger
	rate density of BBHs (top panels) and cosmic BBH merger rate density evolution (bottom
	panels) as obtained from the Model Universe. To obtain the posteriors of $\fymc$ and
	$\fobin$, the differential merger rate densities corresponding to the
	`low-Z', `moderate-Z', and `high-Z' cosmic metallicity evolutions \citep{Chruslinska_2019}
	are considered together (with equal weights), as shown in Fig.~\ref{fig:diffrate_pure2}.
	In this example, fourth moments of the differential rates are
	applied.}
\label{fig:2ch_ace3_ord4_2}
\end{figure*}

Fig.~\ref{fig:diffrate_pure2} shows the resulting $d\ratei/dX$s (100\% universes) 
with the increased error bars. Fig.~\ref{fig:2ch_ace3_ord4_2} shows the
resulting $d\rate/d\mone$, $d\rate/dq$, and $\rz(\zevnt)$ from the
two-channel universe (for $p=4$ and $\ace=3$ IB evolution). The corresponding
total merger rates, merger efficiencies, $\langle\fymc\rangle$,
and $\langle\fobin\rangle$ are quoted in Tables~\ref{tab_rates} and \ref{tab_vals}.
Overall, Fig.~\ref{fig:2ch_ace3_ord4_2} exhibits similarly good agreement
with the GWTC-2 rates as in the previous figures with similar estimated
values of $\langle\fymc\rangle$ and $\langle\fobin\rangle$.
It is, however, important to stress that the error analysis,
as presented above, is incomplete and serves only as a demonstration.
To incorporate uncertainties in a more complete manner,
additional astrophysical sources of uncertainties, \eg, variations of the SFH,
alternative cosmic metallicity evolutions, varied binary evolution
physics (\eg, $\ace$), wider ranges of cluster structure and initial condition
\citep[\eg,][]{Antonini_2020b,Rafelski_2012,Fishbach_2021,Bavera_2020,Gallegos_2021,DiCarlo_2020,Rizzuto_2021}
needs to be considered.
The above exercise demonstrates that
the present data-driven approach can naturally include outcomes from population syntheses
with any set of model assumptions and thus, in principle, can simultaneously incorporate multiple
astrophysical uncertainties.

\section{Summary and concluding remarks}\label{summary}

This study attempts to combine two widely investigated channels of stellar-mass BBH mergers
in the Universe, namely, dynamical interactions involving stellar-mass BHs inside YMCs evolving into medium
mass OCs (the YMC/OC channel) and evolution of isolated massive binaries in the
field (the isolated-binary or IB channel). Cosmological population syntheses
are performed with (hypothetical) universes comprising only model YMC/OCs or IBs
(Secs.~\ref{ymcmodel}, \ref{isomodel}),
taking into account $\Lambda$CDM background cosmology and observation-based
cosmic star-formation and metallicity evolutions
(Sec.~\ref{popsynth}; Figs.~\ref{fig:diffrate_pure}, \ref{fig:rz_pure}).
The resulting present-day differential intrinsic BBH merger rate density
from both universes are then linearly combined assuming constant effective values of 
YMC formation efficiency, $\fymc$, and OB-star binary fraction, $\fobin$,
throughout the cosmic history. The quantities $\fymc$ and $\fobin$ are
then estimated, based on present-day differential intrinsic BBH merger rate density
from GWTC-2, by applying a two-stage linear Bayesian regression involving
moments of the rate distributions (Sec.~\ref{res}).

The resulting Model Universe, combined, present-day differential BBH merger
rate density and also the cosmic evolution of the combined, total BBH merger rate
density agree well with those from GWTC-2 (Sec.~\ref{univ_2ch}; Figs.~\ref{fig:diffrate_2ch_ace1},
\ref{fig:rz_2ch_ace1}, \ref{fig:diffrate_2ch_ace3}, \ref{fig:rz_2ch_ace3}).
The agreements from the two-channel universe are better and more complete than those
from the one-channel universes (Sec.~\ref{univ_1ch}; Figs.~\ref{fig:rates_1ch_ymc},
\ref{fig:rates_1ch_iso}) where BBH mergers are assumed to be produced
via either YMC/OC dynamics or IB evolution.
The estimated $\langle\fobin\rangle\gtrsim90$\%
(see Table~\ref{tab_vals}) is in agreement with the observed high binary fraction among OB stars.
The estimated $\langle\fymc\rangle\sim10^{-2}$ is consistent with cluster
formation efficiencies from recent cosmological simulations.

The physical interpretation of $\fymc$ and $\fobin$ should, however, be taken
with caution. This is especially so for $\fymc$: cluster formation efficiency is
itself subject to varied interpretations (\eg, \cite{Baumgardt_2013,Longmore_2014,Banerjee_2018b}).
From methodological point of view, $\fymc$ and $\fobin$ simply serve like
`branching ratio' or `mixing fraction' \citep{Zevin_2020,Bouffanais_2021},
determining the relative contributions from the two merger channels.

The present results suggest that despite significant BBH-merger contributions from
dynamics in YMCs and OCs at low redshifts, high-redshift ($\zevnt\gtrsim1$) behaviour of  
the BBH merger rate density is still determined by the physics of binary evolution
(Sec.~\ref{univ_2ch}). Hence, future GW detectors with increased visibility horizons,
\eg, LVK {\tt A+} and {\tt A++} upgrades, {\tt Voyager}, {\tt Einstein Telescope},
{\tt Cosmic Explorer} will potentially be able to provide information 
regarding the physical processes in massive-star binaries that drive
compact binary mergers from them. Similar conclusion is drawn in recent independent
studies involving similar binary population synthesis in a one-channel
universe (\eg, \cite{Baibhav_2019,Santoliquido_2020}).

Note that all the estimates and hence the conclusions in this study are subject
to the specifics of the YMC/OC- and IB-evolutionary models (Secs.~\ref{ymcmodel},\ref{isomodel}).
Especially, BBH (and other compact-binary) mergers from IB evolution
is sensitive to crucial binary-evolution ingredients such as models of
tidal interaction, mass transfer, and CE evolution \citep{Giacobbo_2018,Banerjee_2020,Marchant_2021}.
The direct N-body evolutionary models of YMC/OCs treat all Newtonian
and PN interactions explicitly, member-by-member, and without any
symmetry assumptions or modelling them \citep{Aarseth_2012}.
Also, the vast majority of the YMC/OC BBH mergers are dynamically
assembled inside the clusters and hence they do not depend explicitly on
binary-evolution physics.
However, the same $\bse$ that is used for IB evolutions also goes into the
stellar- and binary-evolution modelling during the N-body integration,
shaping the mass distribution of the BHs retained in the cluster (which BHs, eventually,
participate in dynamical pairings). The BH mass distribution depends
on $\bse$'s modelling of star-star and star-BH mergers and also
the ingredients of binary evolution modelling (tidal interaction,
mass transfer, CE evolution) that drive these events
\citep{Spera_2019,Banerjee_2020,Banerjee_2020c}.

The present study is a proof-of-concept demonstration utilizing computations
of model YMC/OCs and IBs. It demonstrates a simplistic linear Bayesian regression
chain involving only raw moments, which statistics are biased quantities.
This will be improved in a future work by incorporating central moments
and/or moments around multiple axes.
While all the analyses and comparisons in this work are done based
on an underlying or intrinsic population model of the LVK (their power law + peak model),
it is important and more model-independent to compare directly with
the posterior samples of the event parameters from GWTC-2 (and future GW-event catalogues)
\citep{Mandel_2019,Perna_2019,Bouffanais_2021}. The analysis will also benefit
by refining the metallicity coverage and expanding the range of $\ace$
of the IB-evolutionary models \citep[\eg,][]{Wong_2021,Broekgaarden_2021}.

In the present demonstration, only two BBH merger channels are considered.
Additional merger channels
and additional types of compact-binary mergers (\ie, NS-BH and BNS mergers)
can be incorporated via straightforward extensions.
Other widely explored channels to consider
\footnote{In principle, any channel whose model provides mergers
with known properties and delay times as functions of
properties of a parent stellar population can be
included in the analysis.}
are chemically-homogeneous binary evolution,
many-body dynamics in GCs, low mass young clusters, and nuclear clusters,
few-body dynamics in field hierarchical systems and AGN
gas discs, pairing of BHs derived from Population-III stars
\citep[\eg,][]{duBuisson_2020,Kremer_2020,Kamlah_2021,Rastello_2021,Antonini_2016,Antonini_2017,Fragione_2020,Secunda_2019,Tanikawa_2021,Ziegler_2021}.
Such range of channels would also help filling up the more extreme regions of
the differential rate distributions (\eg, those with $\mone$ in the PSN gap
and $q\lesssim0.3$).

\begin{acknowledgments}
The author (SB) is thankful to the anonymous referee for constructive criticisms
which have helped to improve the work and the presentation.
SB acknowledges support from the Deutsche Forschungsgemeinschaft (DFG; German Research Foundation)
through the individual research grant ``The dynamics of stellar-mass black holes in
dense stellar systems and their role in gravitational-wave generation'' (BA 4281/6-1; PI: S. Banerjee).
SB acknowledges the generous support and efficient system maintenance of the
computing teams at the AIfA and HISKP.
This work has been benefited by discussions with Chris Belczynski, Mirek Giersz, 
Floor Broekgaarden,
Mark Gieles, Fabio Antonini, Silvia Toonen, Albrecht Kamlah, Rainer Spurzem, 
Manuel Arca Sedda, Peter Berczik,
Giacomo Fragione, Kyle Kremer, Kaila Nathaniel, and Philipp Podsiadlowski.
\end{acknowledgments}

\bibliography{bibliography/biblio}


\end{document}